\documentclass[twocolumn,showpacs,preprintnumbers,amsmath,amssymb]{revtex4}
\usepackage{CJK}

\usepackage{graphicx}
\usepackage{dcolumn}
\usepackage{bm}
\usepackage{tabularx}

\begin{document}
\begin{CJK*}{GBK}{song}

\title{Global analysis of isospin dependent microscopic nucleon-nucleus optical potential in Dirac Bruckner Hartree-Fock approach}

\author{Ruirui Xu }\thanks{xuruirui@ciae.ac.cn}\affiliation{China Institute of Atomic Energy, P.O. Box 275(41), Beijing 102413, China}
\author{Zhongyu Ma }\thanks{mazy12@ciae.ac.cn}\affiliation{China Institute of Atomic Energy, P.O. Box 275(41), Beijing 102413, China}

\author{E. N. E. van Dalen}
\affiliation{Institut f\"{u}r Theoretische Physik, Universit$\ddot{a}$t T$\ddot{u}$bingen, Auf der Morgenstelle 14, D-72076 T$\ddot{u}$bingen, Germany}
\author{H. M\"{u}ther}\thanks{herbert.muether@uni-tuebingen.de}\affiliation{Institut f\"{u}r Theoretische Physik, Universit$\ddot{a}$t T$\ddot{u}$bingen, Auf der Morgenstelle 14, D-72076 T$\ddot{u}$bingen, Germany}

\author{Yue Zhang }
\affiliation{China Institute of Atomic Energy, P.O. Box 275(41), Beijing 102413, China} %
\author{Yuan Tian }
\affiliation{China Institute of Atomic Energy, P.O. Box 275(41), Beijing 102413, China} %

\date{\today}

\begin{abstract}
\noindent\textbf{Background:} For the study of exotic nuclei it is
important to have an optical model potential, which is reliable not
only for stable nuclei but can also be extrapolated to nuclear
systems with exotic numbers of protons and neutrons. An effective way to
obtain such a potential is to develop a microscopic optical
potential (MOP) based on a fundamental theory with a minimal number
of free parameters, which are adjusted to describe stable nuclei all
over the nuclide chart.

\noindent\textbf{Purpose:} The choice adopted in the present work is
to develop the MOP within a relativistic scheme which provides a
natural and consistent relation between the spin-orbit part and the
central part of the potential. The Dirac-Brueckner-Hartree-Fock
(DBHF) approach provides such a microscopic relativistic scheme,
which is based on a realistic nucleon-nucleon interaction and
reproduce the saturation properties of symmetric nuclear matter
without any adjustable parameter. Its solution using the projection
technique within the subtracted T-matrix (STM) representation
provides a reliable extension to asymmetric nuclear matter, which is
important to describe the features of the isospin asymmetric nuclei.
Therefore, the present work aims to perform a global analysis of the
isospin-dependent nucleon-nucleus MOP based on the DBHF calculation
in symmetric and asymmetric nuclear matter.

\noindent\textbf{Methods:} The DBHF is used to evaluate the
relativistic structure of the nucleon self-energies in nuclear
matter at various densities and asymmetries. The Schr\"{o}dinger
equivalent potentials of finite nuclei are derived from these Dirac
components by a local density approximation (LDA). The density
distributions of finite nuclei are taken from the
Hartree-Fock-Bogoliubov (HFB) approach with Gogny D1S force. An
improved LDA approach (ILDA) is employed to get a better prediction
of the scattering observables. A $\chi^{2}$ assessment system based
on the global simulated annealing algorithm (GSA) is developed to
optimize the very few free components in this study.

\noindent\textbf{Results:} The nucleon-nucleus scattering
calculations are carried out for a broad spectrum $n$ and $p$
scattering experiments below 200 MeV with targets ranging from
$^{12}$C to $^{208}$Pb. The scattering observables including the
neutron total cross section, proton reaction cross section, elastic
scattering angular distribution, analyzing power and spin rotation
are evaluated and compared with the experimental data, as well as
with results derived from the widely used phenomenological
Koning-Delaroche (KD) global potential.

\noindent\textbf{Conclusions:} Results with the present relativistic
MOP satisfactorily reproduce the $n, p$ + $A$ scattering observables
over a broad mass range and a large energy region only with the
free range factor $t$ in ILDA and minor adjustment to the scalar and vector potentials around the low density region,
and the relevant potentials are physically well-behaved. The overall agreement indicates that the
present MOP may have predictive power for unstable nuclei, as well
as the nuclei beyond the line of $\beta$ stability.

\end{abstract}

\maketitle

\section{Introduction}

The nuclear reaction of unstable nuclei is of high interest in the
contemporary fundamental physics as well as applied nuclear physics.
The corresponding information from experimental data is always too
sparse or totally lacking. Therefore, during the past few decades,
considerable effort has been made to explore the nuclear reaction
based on a fundamental microscopic nuclear theory so as to obtain
more reliable prediction.

The optical model is the crucial component in the nuclear reaction
study, mainly because it determines the cross section for nuclear
scattering and compound nuclear formation in the initial stage of a
reaction and supplies the transmission coefficients for branching
into the various final states \cite{M.T.Pigni2011}. Many observables
such as the elastic scattering angular distribution, analyzing
power, spin rotation function and so on can be derived through the
optical model. Therefore, the most important criteria to assess a
microscopic optical potential (MOP) is that it can reproduce the
existing experimental data of these observables as accurate as
possible, and make a reliable prediction without experimental
guidance. Furthermore, the MOP is more appealing when it is
established on better theoretical grounds with a small number of
free parameters.

In the direct evaluation for MOP of finite nuclei an attempt is made to evaluate
the scattering and absorption processes using a many-body theory for the target
nuclei which goes beyond mean field theory  and incorporates e.g. the effects of
particle-vibration couplings. These studies typically employ an
effective nucleon-nucleon ($NN$) interaction (e.g. Skyrme interaction
or Gogny interaction). Recently, several investigations for finite
nuclei have been reported \cite{G.Blanchon,Li-Gang.Cao}, however, it
seems still infeasible to derive the MOP for all nuclei,
which are of interest e.g. in the field of applications of nuclear
physics. In addition the specification of a sound nuclear structure
for targets, especially for exotic nuclei remains to be progressed.
Using phenomenological $NN$ interactions such calculations are designed
to derive the nuclear structure and the MOP from the same interaction model.
A drawback of this scheme is phenomenological and fitted
to describe these data.

On the other hand, various attempts have been made to derive the MOP
from a realistic model of the $NN$ interaction, which means an interaction
designed and fitted to describe the $NN$ scattering data. Such studies often
use the system of nuclear matter to determine the effects of correlations and
evaluate the medium dependence of the resulting effective interaction for nuclear
matter. The nuclear matter results are then used in various kinds of local density approximation (LDA)
to be applicable for finite nuclei. The review article of Ray et al. discusses various
approximation schemes along this line \cite{ray92}.

Pioneering work along this line has been presented by Mahaux and coworkers \cite{Jeu77}
who evaluated the nucleon self-energy in nuclear matter as a function of density and energy
in a Brueckner Hartree Fock (BHF) approximation and identified the resulting complex single-particle
potential with the MOP for finite nuclei using LDA adopting
nucleon density distributions from the empirical formula or microscopic nuclear structure calculations.
One drawback of this scheme is that typically one has to use an interaction model for the evaluation of the
density profile of the nuclei which is different from the realistic interaction used to calculate the
self-energy in nuclear matter as BHF calculations fail to reproduce the empirical saturation
properties for nuclear matter and finite nuclei.
In a simplified way, some MOPs have been developed by adopting the effective $NN$ interactions (e.g. Skyrme force) in the Hartree-Fock approach
in nuclear matter and LDA for finite nuclei \cite{Y.Xu2014}.

Another handicap of this approach is the fact that this approach only provides the
central part of the MOP, the spin-orbit potential has to be adjusted independently
from the central potential in such a non-relativistic approach.
Nevertheless this scheme has been applied with quite some success by
Jeukenne, Lejeune and Mahaux already in the
1970s \cite{Jeu77} and is still rather popular today.

Also the so-called $g$-folding method developed by Amos et al. \cite{K.Amos2000} is based
on a realistic $NN$ interaction and uses a local density approximation to account
for the medium dependence of the effective interaction. In this case, however, it is the $NN$
interaction, which is evaluated by solving the Bethe-Goldstone equation in nuclear matter and
then employed in a folding calculation to evaluate the MOP for finite nuclei. The $g$-folding
approach has very successfully been applied to reproduce differential cross sections and spin
observables in an energy range from 30 MeV to 300 MeV in several nuclei
without any adjustment of parameters \cite{K.Amos2000,deb01,brown00,P.K.Deb2005,M.Dupuis2006,M.Dupuis2008,A.Lagoyannis2001}.

The Mahaux scheme as well as the "g"-folding method are based on a non-relativistic approach and the energy-dependence of
the MOP originates from the energy-dependence of the effective interaction $g$ calculated
for nuclear matter in the BHF approximation.

An alternative approach is based on the
Dirac phenomenology as it has been introduced by Walecka and coworkers \cite{walecka}. Within this Dirac phenomenology the nucleon self-energy contains a large and attractive
component, which transforms like a scalar under a Lorentz transformation compensated to a large
extent by a repulsive Lorentz vector component. If one reduces the corresponding Dirac equation
for the nucleon in the nuclear matter medium to a  non-relativistic Schr\"{o}dinger equation, one
obtains a Schr\"{o}dinger equivalent potential with a central potential which is energy dependent
and a strong spin-orbit term.

An application of this Dirac phenomenology to describe the optical model potential has been
presented by Cooper et al. \cite{cooper}. They developed a phenomenological parametrization of
the real and imaginary parts of the scalar and vector potentials. Fitting the corresponding parameters,
which depend on energy and mass number of the target nucleus, they obtain a very good global fit
of the optical model potential.

A comparison of the rather successful but very different approaches to a global optical potential,
the $g$-folding method and Dirac phenomenology, has been made by Deb et al. \cite{P.K.Deb2005}.
They evaluated differential cross sections and spin observables for nucleon nucleus scattering
on five different targets ranging from $^{12}$C to $^{208}$Pb at energies of 65 and 200 MeV
using both approaches and conclude that the results are of similar quality.

It is one aim  of the Dirac Brueckner Hartree Fock (DBHF) approach, to combine the features of a
realistic $NN$ interaction and its dependence on the medium, as they are contained in the
Mahaux approach and the "g"-folding model with those of the Dirac phenomenology \cite{RRXU2012}.
This approach is founded on a realistic $NN$ interaction and the treatment of
nuclear correlations and the medium dependence of the effective $NN$ interaction is done in straight
analogy to the Mahaux approach and the
$g$-folding method. The DBHF approach, however, keeps track of the relativistic structure of the
nucleon self-energy and therefore one can determine the real and imaginary part of the
scalar and vector component of the nucleon self-energy in nuclear matter as a function
of momentum, density and energy. These components are then used to evaluate the
corresponding MOP using LDA for these components of the self-energy
in straight analogy to the Mahaux scheme.

In this way, the spin-orbit potential arises naturally from the coherent sum of the contribution from the
scalar and vector potentials in this relativistic scheme, and
the saturation properties of symmetric nuclear matter are reproduced
in the relativistic Dirac Brueckner-Hartree-Fock (DBHF) approach, while
three-nucleon forces have to be introduced to obtain a corresponding
result within the non-relativistic BHF approximation \cite{dalrev,lipok,Z.H.Li2008}.
Therefore it seems rather attractive to determine a microscopic
optical model based on the DBHF approach as finally one may be able to describe
the ground-state properties of nuclei and the MOP within the same theoretical framework.

For a long period it has been a challenge in theoretical nuclear
physics to solve the Brueckner-Hartree-Fock in the relativistic way,
especially for isospin asymmetric nuclear matter. In recent years
substantial progress has been obtained using a so-called subtracted
T-matrix (STM) representation in the projection technique to solve
DBHF strictly in the symmetric and asymmetric nuclear matter
\cite{Dalen04}.

In a preliminary study we have explored the isospin dependent
relativistic microscopic optical potential adopting the
self-energies from this DBHF calculation in \cite{RRXU2012}. This
MOP has been verified by satisfactorily reproducing the neutron and
proton scattering data from $^{27}$Al. In this work, a systematic
investigation for this MOP is performed in a large range of nuclei.
The microscopic radial nucleon density of finite nuclei based on the
Hartree-Fock-Bogoliubov (HFB) calculation are adopted in this
calculation instead of the previous empirical values. Meanwhile, a
$\chi^{2}$ assessment system based on the global simulated annealing
algorithm (GSA) is specially designed to optimize the free factors
and give an overall estimation on the performance of this MOP.

The paper is composed as follows. In Section II, the general
formalism of DBHF is briefly introduced. The isospin dependent
relativistic MOP of finite nuclei are built in Section III through
combining the self-energies and the microscopic radial nucleon
density by the improved local density approximation (ILDA)
\cite{Jeu77}. The global analysis of nuclear scattering is carried
out in Section IV for neutron and proton scattering and induced
reactions on $^{12}$C - $^{208}$Pb and the calculated results are
compared with the calculated results with the widely-used
phenomenological Koning-Delaroche (KD) global optical potential
\cite{Kon03} and the experimental data of various scattering
quantities. Finally, the overall discussion is summarized in Section
V.

\section{SELF-ENERGY IN NUCLEAR MATTER}

Realistic $NN$ interactions contain strong short-range and tensor
components. Therefore it is necessary to account for the
corresponding correlations between the interacting nucleons. In the
relativistic Brueckner-Hartree-Fock approach this is achieved by
considering the equation for two interacting nucleons in nuclear
matter. This leads to the ladder approximation of the relativistic
Bethe-Salpeter (BS) equation \cite{Gross99,Dalen04},
\begin{eqnarray}
     T = V + i \int VQGGT, \label{eq1}
\end{eqnarray}
where $T$ is the nucleon-nucleon interaction matrix in the nuclear
medium and $V$ is the bare $NN$ interaction, respectively. The Pauli
exclusion principle is included by the $Q$ operator and the
in-medium nucleon propagation of the nucleons is described by the
Green's function $G$, which fulfills the Dyson equation,
\begin{eqnarray}
     G = G_0 + G_0 \Sigma G. \label{eq2}
\end{eqnarray}
$G_{0}$ denotes the free nucleon propagator, and the self-energy
term $\Sigma$ is defined in first order of the effective interaction
$T$ through the following standard Hartree-Fock equation
\begin{eqnarray}
     \Sigma = - i \int_F (Tr[GT]-GT).  \label{eq3}
\end{eqnarray}
Note  the self-energy contains the direct and exchange terms at the
same time, and that the momentum integration considers all nucleon
states within the Fermi sea represented by F in Eq. \ref{eq3}. Because Eqs. \ref{eq1}-\ref{eq3} are
strongly coupled, they have to be solved iteratively until
convergence is reached.

Generally, the Lorentz structure of the relativistic self-energy
$\Sigma$ can be expressed as \cite{Ma88},
\begin{eqnarray}
     \Sigma^m(k,k_F,\beta) = \Sigma^m_s(k,k_F,\beta) - \gamma_0\Sigma_0^m(k,k_F,\beta) \label{eq5}\\
     \nonumber + \bm{\gamma}\cdot
     \textbf{k}\Sigma_v^m(k,k_F,\beta)~.
\end{eqnarray}
In this equation, $\Sigma_s$ is the scalar part of self-energy, $\Sigma_0$ and
$\Sigma_v$ denote the time-like and space-like terms of the vector part, respectively. The superscript $m$ is used
to sign the proton and neutron since they should be distinguished in
isospin asymmetric nuclear matter. Note that these components of the
self-energy are functions of the nucleon momentum ($k$), density or
Fermi momentum ($k_F$), and asymmetry parameter $\beta$ =
($\rho_n$-$\rho_p$)/$\rho$, where $\rho_n$, $\rho_p$ and $\rho$
indicate the neutron, proton and total densities in nuclear matter,
respectively.

Details of such DBHF calculations and the method to extract these
Dirac components using the subtracted T-matrix (STM) representation
are described in \cite{Dalen04,Dalen07,Dalen10}. The self-energies
used in the present study are determined using the Bonn-B potential
\cite{Machxxx} for the bare $NN$ interaction and solving the DBHF
equations for isospin asymmetric nuclear matter with various
densities and isospin asymmetries.

\section{RELATIVISTIC MICROSCOPIC OPTICAL POTENTIAL IN FINITE NUCLEI}
In the relativistic scheme, the wave function of an incident
particle described in terms of a Dirac spinor $\Psi$  is obtained by
the solution of the corresponding Dirac equation,
\begin{eqnarray}
      \left[\vec{\alpha}\cdot\vec{p}+\gamma_0(\textit{M}+U_s^m)+U_0^m \right]
      \Psi^m=\varepsilon\Psi^m~ , \label{eq10}
\end{eqnarray}
where $U_s^m$ and $U_0^m$ are the scalar and vector components of the scattering potential
\begin{eqnarray}
      U_s^m=\frac{\Sigma_s^m-\Sigma_v^m\textit{M}}{1+\Sigma_v^m},~~U_0^m=\frac{-\Sigma_0^m+\varepsilon\Sigma_v^m}{1+\Sigma_v^m}\,,~~ \label{eq11}
\end{eqnarray}
and $\varepsilon$ = $\textit{E}+\textit{M}$ is the single particle
energy, $E$ is the kinetic energy of the nucleon in the free space
and $M$ indicates the mass of the nucleon.

In order to calculate the scattering observables of finite nuclei
this Dirac equation is typically reduced to a Schr\"{o}dinger type
equation  by eliminating the lower components of the Dirac spinor in
a standard way. The equation for the upper components of the wave
function is transformed into:
\begin{eqnarray}
\left[-\frac{\nabla^2}{2{\varepsilon}}+V^m_{cent}+V^m_{s.o.}(r)\vec{\sigma}\cdot\vec{\textbf{\L}}+V^m_{Darwin}(r)
\right]\varphi(\textbf{r}) \\
      \nonumber =\frac{\varepsilon^2-\textit{M}^2}{2\varepsilon}\varphi(\textbf{r}), \label{eq13}
\end{eqnarray}
where $V^m_{cent}$, $V^m_{s.o.}$ and $V^m_{Darwin}$ represent
the Schr\"{o}dinger equivalent central, spin-orbit
and Darwin potentials, respectively.
The potentials in Eq. \ref{eq13} are obtained from the scalar $U_s$ and vector $U_0$ potentials as
\begin{eqnarray}
      \nonumber V^m_{cent}=\frac{M}{\varepsilon}U_s^m+U_0^m+\frac{1}{2\varepsilon}[U_s^{m2}-(U_0^m+V_c)^2], \label{eq14}
\end{eqnarray}
\begin{eqnarray}
      V_{s.o.}^m=-\frac{1}{2{\varepsilon}rD^m(r)}\frac{dD^m(r)}{dr}, \label{eq15}
\end{eqnarray}
\begin{eqnarray}
     \nonumber V_{Darwin}^m=\frac{3}{8{\varepsilon}D^m(r)}[\frac{dD^m(r)}{dr}]^2-\frac{1}{2{\varepsilon}rD^m(r)}\frac{dD^m}{dr} \\
     \nonumber -\frac{1}{4{\varepsilon}D^m(r)}\frac{d^2D^m(r)}{d^2r},\label{eq16}
\end{eqnarray}
where $V_c$ is the Coulomb potential for a charged particle and $D$
denotes a  quantity defined as
\begin{eqnarray}
       D^m(r)=M+\varepsilon+U_s^m(r)-U_0^m(r)-V_c. \label{eq17}
\end{eqnarray}

The radial potentials in finite nuclei, namely $V^m_{cent}$,
$V^m_{s.o.}$, and $V^m_{Darwin}$ in Eqs. (\ref{eq14})-(\ref{eq17}),
can be associated with the scalar $U_s$ and vector $U_0$ in nuclear
matter through the local density approximation (LDA) using the local
nucleon density $\rho(r)$ for the nucleus considered. In this work,
a finite range correction in Gaussian form is adopted in LDA to
further remedy the potentials to obtain the better prediction of the
scattering experimental data, that is the so-called improved local
density approximation (ILDA),
\begin{eqnarray}
       U_{ILDA}(r,E)&=&(t\sqrt{\pi})^{-3}\times \label{eq18}\\
     && \int{U_{LDA}(r',E)exp(-|\vec{r}-\vec{r'}|^2/t^2)d^{3}r'}~, \nonumber
\end{eqnarray}
where $t$ is an effective range parameter of the potential $U_{LDA}$
in normal LDA approach at radius $r'$. It is included to account for
a finite-range correction of the nucleon-nucleon interaction, which
is not incorporated in the DBHF calculation. They modify the radial
distribution of V$_{cent}$ while keep its volume integral constant.
The potential $U_{LDA}$ is related to the $U_s$ and $U_0$ in nuclear
matter by
\begin{eqnarray}
      U_{LDA}(r,E)=U_{NM}(k,E,\rho(r),\beta(r)), \label{eq9}
\end{eqnarray}
and $U_{NM}$ represent the corresponding potential in nuclear matter
using the isospin asymmetry $\beta$ of the target nucleus and the
momentum $k$ and energy $E$ of the incoming nucleon.
In our present studies we adopt the radial nucleon density,
$\rho(r)$, from the Hartree-Fock-Bogoliubov (HFB) approach with
Gogny D1S force \cite{Hilaire2007}, instead of the empirical values
by the Negele's formula \cite{Negele70}, which has been employed for
our pilot study \cite{RRXU2012}.

\begin{figure}[htbp] \centerline{\includegraphics[width = 3.5in]{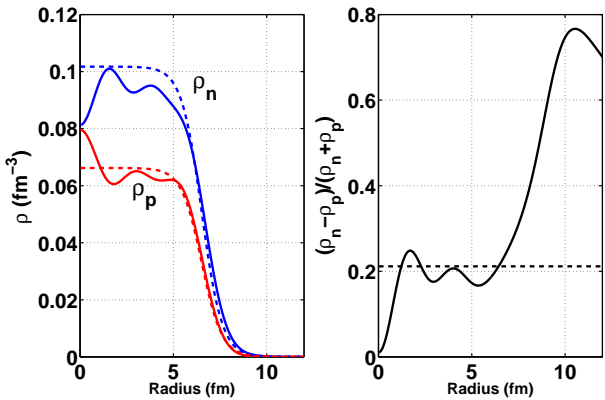}}
\caption{(color online) The proton and neutron radial densities for $^{208}$Pb. The solid and dashed lines indicate the calculated
results from HFB and the Negele's empirical formula, respectively.}
\label{fig1}
\end{figure}

In Fig. \ref{fig1}, we compare the radial density and asymmetry
distributions for $^{208}$Pb as derived from HFB approach and the
empirical formula. The radial densities obtained in the HFB approach
show oscillations in the interior of the nucleus, which reflect the
structure of the single-particle wave functions. Note, however, that
the oscillations are smoothed out to a large extend in the ILDA
potentials by the finite range correction of Eq. \ref{eq18}. Also
note the enhancement of the neutron density in the surface of the
nucleus in the microscopic calculation. This neutron skin leads to
large isospin asymmetries as presented in the left panel of Fig. \ref{fig1}.

In addition, as for this microscopic optical potential, the
applicability of the theory should be indicated that the formalism
does not include the coupling to giant resonances (10 MeV to 30 MeV) and
the compound nucleus formation ($<$ 10 MeV), which has been
discussed before and stressed in e.g. \cite{K.Amos2000}. At the low
energies, we include the compound nuclear contribution to the
elastic differential cross sections by the Hauser-Feshbach statistic
theory through the optical model code APMN \cite{Shen02}, which
employs the Hauser-Feshbach model to determine the contributions
from the compound nuclear elastic scattering by concerning six
competing single-particle emission reactions, including neutron,
proton, deuteron, tritium, $\alpha$, and $^3$He. The formation of
giant resonances will further discussed in other future work.

\section{GLOBAL ANALYSIS OF NUCLEON-NUCLEI RMOP}
\subsection{Dirac potentials in the density region relevant for finite nuclei}
It is the aim of this work to generate an optical potential,
which has a microscopic basis and reproduces bulk features of
nucleon-nucleus scattering for nuclei across the whole nuclide
chart. The microscopic basis of these calculations originates from
the real and imaginary parts of the Dirac components, $U_s^m$ and
$U_0^m$, of the nucleon self-energies calculated in the DBHF
approach for symmetric and asymmetric nuclear matter. Such DBHF
calculations, however, yield reliable results only for densities
$\rho >$ 0.08 fm$^{-3}$. The procedure to derive self-consistent
DBHF results does typically not converge at lower densities. This
reflects the situation that homogeneous nuclear matter is unstable
at such low densities with respect to the formation of an
inhomogeneous density profile containing nuclear clusters. In
particular the solution of the $T$-matrix of Eq. \ref{eq1} yields
bound states in the deuteron channel.

For the derivation of the optical model potential for finite nuclei,
however, we also need results at densities $\rho<$ 0.08 fm$^{-3}$.
Therefore we have to extrapolate the results to these low densities
with the natural constraint that the Dirac potentials $U_s^m$ and
$U_0^m$ vanish at $\rho = 0.$ A linear interpolation of these Dirac
potentials between $\rho =$ 0.08 fm$^{-3}$ and $\rho$ = 0 might be
too simple. Therefore we have introduced auxiliary mesh-points at
$\rho =$ 0.04 fm$^{-3}$ for the real parts and at $\rho =$ 0.04 and
0.06 fm$^{-3}$ for the imaginary parts. The values of the linear
interpolation at these auxiliary mesh-points is then enhanced by a
factor $f_1$ for the real part and a factor $f_2$ for the imaginary
part. Typical value for these enhancement factors ($f_1,f_2$) are
(0.86, 1.14) (see description of fitting procedure below).

Since the evaluation of the spin-orbit and Darwin potential (see
Eq. \ref{eq15}) requires the calculation of derivatives of the
Dirac potentials with respect to their density dependence, it is
favorable to determine a simple interpolation scheme, which is valid
for all densities entering into the calculation. Therefore we have
chosen to represent the density-dependence of the Dirac potentials
in terms of a polynomial fit with a polynom of degree 5 for the real
part and a polynom of degree 7 for the imaginary part. The
coefficients of these polynomials are determined to fit the results
at all calculated densities, including the auxiliary mesh points
mentioned above.

\begin{figure*}[htbp]
\centerline{\includegraphics[width = 5.0in]{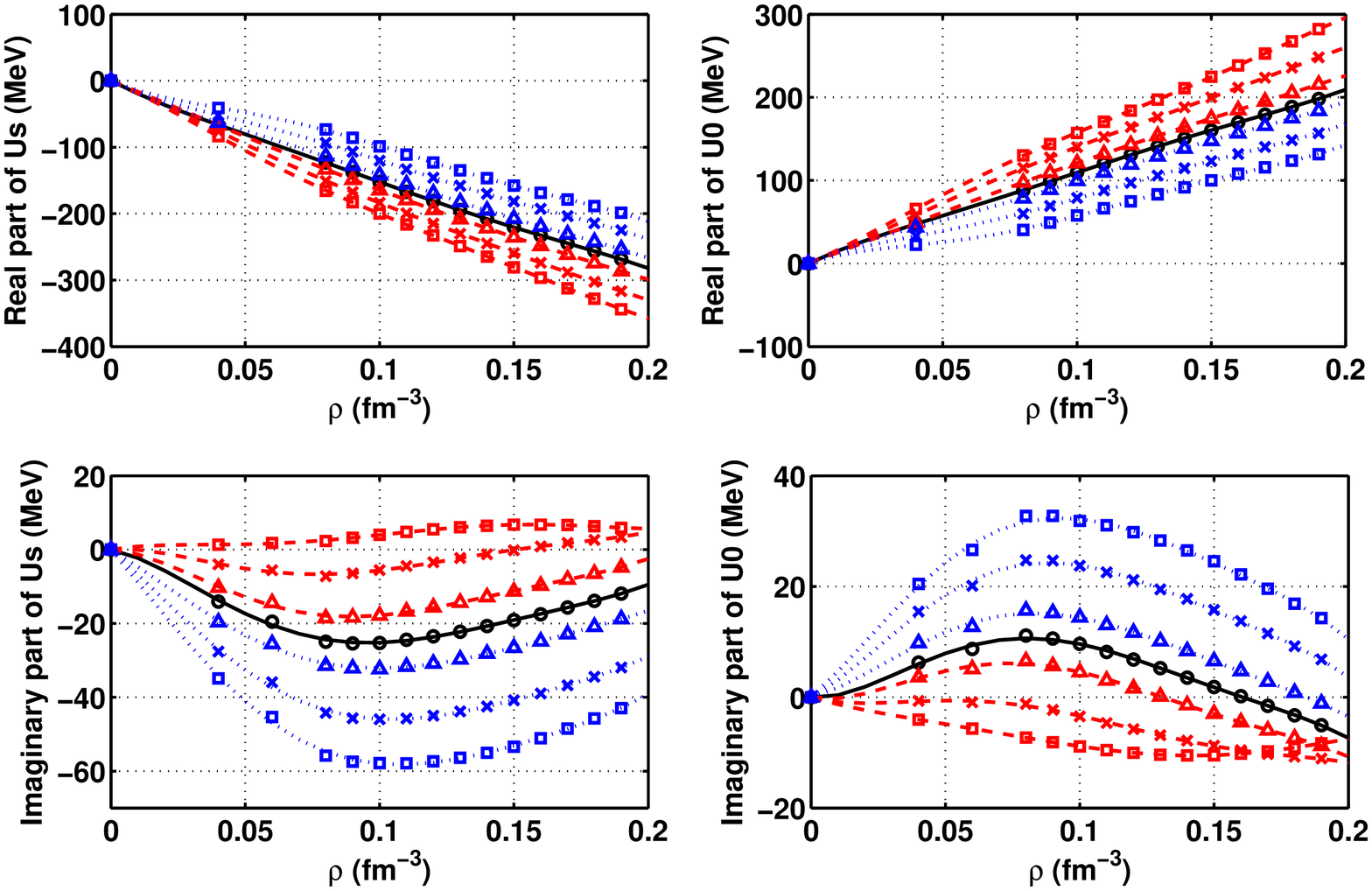}}
\caption{ (color online) Example for the real and imaginary part of the scalar ($U_s$) and
vector ($U_0$) components of the Dirac potential as a function of density for nucleons with
an incident particle energy of 90 MeV. The symbols circle,
triangle, x-mark and square represent the calculated (adjusted) values for isospin asymmetries
$\beta =$ 0.0, 0.2, 0.6 and 1.0, respectively. The connecting solid line shows the polynomial
 interpolation in
the case of symmetric matter, while the dashed and dotted lines visualize the corresponding
interpolations for the neutron- and proton-potentials, respectively.}
\label{fig2}
\end{figure*}

As an example we present in Fig. \ref{fig2} values for the real and
imaginary parts of the Dirac potentials $U_s^m$ and $U_0^m$ for
nucleons with an energy of 90 MeV. The calculated values for
densities ranging between 0.08 and 0.2 fm$^{-3}$, as well as those
for the auxiliary mesh-points are represented by circles, triangles,
x-marks, and squares for isospin asymmetries $\beta =
(\rho_n-\rho_p)/\rho$ of 0.0, 0.2, 0.6 and 1, respectively. The
corresponding polynomial interpolations are visualized in terms of a
solid line for isospin symmetric nuclear matter ($\beta = 0$), while
the dashed line shows the interpolation for the neutron and the
dotted line for the proton potentials at $\beta > 0$. $U_s^m$ and
$U_0^m$ depend in a very strong but also rather smooth way on the
neutron-proton asymmetry parameter $\beta$ at all nuclear densities,
as can be seen from Fig. \ref{fig2}.

Meanwhile, the potentials at a given density between 0.0 and 0.07
fm$^{-3}$ are established with the requirement to maintain the
natural tendency of microscopic DBHF curves. More details about this
system will be introduced later in this section.

\subsection{RMOP optimization}
As mentioned above, one of the main criteria to evaluate a good
optical model potential is that it can well reproduce as many of the
measured scattering observables as possible. Thus, how to optimize
the present microscopic potentials to predict the available
experimental data sets is an important issue. Optimization
procedures have widely been discussed in the literature of the
optical model \cite{Kon03,Perey63}. Most of them obtain the
parameters through minimizing a certain $\chi^{2}$ value, given for
example by
\begin{eqnarray}
     {\chi}^2 = \sum\limits_{i=1}^{P}\left(\frac{\sigma_{i}^{cal}-\sigma_{i}^{exp}}{\sigma_{i}^{exp}}\right)^2, \label{eq12}
\end{eqnarray}
where $\sigma_{i}^{exp}$ is the $i$th experimental point,
$\sigma_{i}^{cal}$ is the corresponding calculated result, and $P$
indicates the total number of experimental data in our
consideration. In this study, the effective range factor $t$ related
to the Schr\"{o}dinger equivalent potentials in Eq. \ref{eq18}, as
well the Dirac potentials at density $\rho =$ 0.04 and 0.06
fm$^{-3}$, remain to be determined in the optimization process. A
$\chi^{2}$ assessment system is specially designed to fulfill this
optimization. We employ the global simulated annealing (GSA) method
based on the Monte Carlo sampling in a predefined region of the free
parameters. The value of $\chi^{2}$/$N$ is adopted as the most
important criteria for this optimization procedure. Here $\chi^{2}$
is defined in Eq. \ref{eq12}, and $N=P-F$ is the number of degrees of
freedom with the total number of experimental data $P$ and the
number of free parameters $F$. It is known, however, that it is not
possible to obtain a 'best-fit' through the numerical optimization
procedure \cite{Kon03} alone, therefore a visual goodness-of-fit
estimation evaluator is also incorporated in our optimization
procedure.

Observables including the angular distribution
(d$\sigma$/d$\Omega$), analyzing power (A$_y$) for nucleon-nucleus
elastic scattering as well as the neutron total cross section
($\sigma_{tot}$) and proton total reaction cross section
($\sigma_{reac}$) et al. have been considered in this optimization
assessment system. In a first step the angular distribution of
elastic scattering, d$\sigma$/d$\Omega$, is taken into account to
optimize the $\chi^{2}$/$N$ between experimental data and our
theoretical calculations. After the 'minimal' $\chi^{2}$/$N$ value
is achieved for d$\sigma$/d$\Omega$, the other observables are then
utilized in the visual goodness estimation for a further improvement
of the MOP.

$^{40}$Ca and $^{208}$Pb, for which a large number of experimental
data of proton and neutron scattering were measured, are both double
magic nuclei and they represent proper examples to cover a good
range from isospin symmetric to asymmetric nuclei, therefore, they
were selected in our MOP optimization procedure. Moreover, after
optimizing the MOP using these data, we further test the potentials
by predicting scattering observables of other nuclei. As a result,
our optimization procedure for proton scattering yields a value of $t$
= 1.35 fm with the value $\chi^{2}/N$ of 0.29, which is a pretty
good result in view of the 1028 points of d$\sigma$/d$\Omega$ in 18
different experimental data sets. Similarly, we get an optimized
$t$ = 1.45 fm for neutron scattering reactions. These values are
comparable to the range parameter $t$ = 1.4 fm, which has been derived
in our pilot study \cite{RRXU2012} focused on the target nucleus
$^{27}$Al.

\subsection{Experimental database}
The present MOP is assessed through a global prediction and analysis
for the main observables of neutron and proton induced scattering
reactions in a large mass region of 12$\leq$A$\leq$ 209 below
incident energy 200 MeV. The most abundant natural isotopes are
considered, and nuclei with even as well as odd mass numbers are
incorporated.

As well known, EXFOR library is a comprehensive database to gather
the nuclear reaction measurements in the world \cite{exf}. The
experimental data adopted in our analysis are all referred in this
library. The details of measured elastic scattering angular distribution are specified in
this paper by the first author and published time, which are shown
in Table \ref{tab1}-\ref{tab2} for neutron induced reaction and
Table \ref{tab6} for proton incident reaction according
to the diversified target nuclei. Other measurements like neutron
total cross section and proton reaction cross section are also
depicted in the same way in figures.

\subsection{Results for neutron scattering}

About 500 sets of elastic scattering angular distributions, 30 sets
of analyzing power angular distributions, and 20 sets of total
neutron cross sections for 32 different targets are involved in this
systematic comparison. The present calculations are compared with
experimental data and the results from the widely-used
Koning-Delaroche (KD) optical potential.

Because nucleon densities of very light nuclei are not described in
a reliable way by means of the Hartree-Fock-Bogoliubov approach, we
take $^{12}$C as the lightest target in this study. Overall, the
predictions of the MOP are in rather good  agreement with the
experimental data as well as the results calculated by KD potential
for such large mass and energy ranges. Meanwhile, it is also
observed that the performance of global KD potential is satisfactory
even beyond its application scope. The discussion is given in the
following subsections in detail.

\begin{table*}
\centering \caption{\label{tab1} The $d\sigma/d\Omega$ database for neutron elastic scattering}
\begin{ruledtabular}
\begin{tabular}{lllllll}
    \toprule
    \textbf{Target} & \textbf{Author(1$^{st}$)} & \textbf{Year} & \textbf{Energy(MeV)} & \textbf{Author(1$^{st}$)} & \textbf{Year} & \textbf{Energy(MeV)} \\
\hline
    6-C-12  & R.O.Lane & 1961  & 1.04,~2.25 & P.Boschung & 1971  & 4.04 \\
          & R.M.White & 1980   & 6.94         & G.Haouat & 1975    & 8.5,~9.0 \\
          & D.W.Glasgow & 1976 & 10.69,~12.49,~13.94 & N.Olsson & 1988    & 17.6,~22.0, \\
          & T.Niizeki & 1990   & 35.0         & J.H.Osborne & 2004 & 65.0,~107.5,~155,~225 \\
          & M.Ibaraki & 2002   & 75.0         & P.Mermod & 2006    & 94.8 \\
\hline
    7-N-14 & J.L.Fowler & 1955 & 1.08,~1.68,~2.07 & F.G.Perey  &  1974 & 4.34,~4.92,~6.01,~7.03,~8.56 \\
          & J.Chardine & 1986 & 7.9,~9.0,~13.5 & D.Schmidt  & 2003 & 10.81,~12.79 \\
          & L.Anli     & 1989 & 14.0,~17.0 & N.Olsson  & 1989 & 21.6 \\
\hline
    8-O-16& L.Drigo & 1976 & 2.56  & I.A.Korzh & 1980 & 5 \\
          & G.Boerker & 1988 & 6.37,~7.51,~9.01,~10.31,~13.61,~14.89 & M.Baba  & 1988 & 14.1 \\
          & L.Anli & 1989 & 17    & J.P.Delaroche & 1986 & 18.0,~26.0 \\
          & N.Olsson & 1989 & 21.6  & P.Mermod & 2006 & 94.8 \\
\hline
    11-Na-23 & W.E.Kinney & 1976 & 0.55,~0.7,~1.0,~1.2,~1.4,~1.6,~1.7,~2.0 & U.Fasoli & 1969 & 1.51,~2.47,~4.04 \\
          & Th.Schweitzer & 1978 & 3.4   & R.E.Coles & 1971 & 5.0 \\
          & F.G.Perey & 1970 & 5.44,~6.37,~7.6,~8.52 & P.Kuijper & 1972 & 14.8 \\
\hline
    12-Mg-24 & D.B.Thomson & 1962 & 3.79  & I.A.Korzh & 1994 & 5.0,~6.0,~7.0 \\
          & W.E.Kinney & 1970 & 7.55,~8.56 & M.Adel-Fawzy & 1985 & 8.0,~9.0,~10.0,~11.0,~12.0 \\
          & A.Virdis & 1981 & 9.76,~14.8 & A.Takahashi & 1987 & 14.1 \\
          & N.Olsson & 1987 & 21.6  &       &  \\
\hline
    13-Al-27 & R.L.Becker  & 1966 & 3.2   & W.E.Kinney & 1970 & 5.44,~6.44,~7.54,~8.56 \\
          & G.Dagge & 1989 & 7.62  & C.S.Whisnant & 1984 & 10.87,~13.88,~16.9 \\
          & M.M.Nagadi & 2003 & 15.4  & J.S.Petler & 1985 & 18,20,~22,25,~26.0 \\
          & A.Bratenahl & 1950 & 84.0    & G.L.Salmon & 1960 & 96.0 \\
          & C.P.Van.Zyl & 1956 & 136.0   &       &  \\
\hline
    14-Si-28 & W.E.Kinney & 1970 & 5.44,~6.37,~6.44,~7.55,~8.56 & C.R.Howell & 1988 & 7.96,~9.95,~11.94,~13.97,~16.92 \\
          & J.Rapaport & 1977 & 11.0,~20.0,~25.0 & R.Alarcon & 1986 & 21.7 \\
          & M.Ibaraki & 2002 & 55.0,~65.0,~75.0 &       &  \\
\hline
    15-P-31 & K.Tsukada & 1961 & 3.5,3.8,4.2,4.5 & J.Martin & 1968 & 5.95 \\
          & J.D.Brandenberge & 1972 & 7.79,9.05 & P.H.Stelson & 1965 & 14.0 \\
          & G.C.Bonazzola & 1965 & 14.2  &       &  \\
\hline
    16-S-32 & F.G.Perey & 1970 & 3,4,~7.05,~7.6,~8.52 & S.Tanaka & 1969 & 5.92 \\
          & C.R.Howell & 1988 & 7.96,~9.95,~11.93,~13.92 & J.D.Brandenberge & 1972 & 9.05 \\
          & A.Virdis & 1981 & 9.76  & J.Rapaport & 1977 & 20.0,~26.0 \\
          & Y.Yamanouti & 1977 & 21.5  & R.Alarcon & 1986 & 21.7 \\
          & J.S.Winfield & 1986 & 30.3,~40.3 &       &  \\
\hline
    19-K-39 & J.H.Towle & 1965 & 1.49,~2.38 & J.D.Reber & 1967 & 2.06,~3.74,~4.33,~6.52,~7.91 \\
          & A.J.Frasca & 1966 & 14.0    &       &  \\
\hline
    20-Ca-40 & J.D.Reber & 1967 & 2.06,~3.29,~5.3,~7.91 & B.Holmqvist & 1969 & 6.09,~7.05 \\
          & W.Tornow & 1982 & 9.91,~11.9,~13.9 & G.M.Honore & 1986 & 16.9 \\
          & R.Alarcon & 1987 & 19.0,~25.5 & J.Rapaport & 1977 & 20.0 \\
          & N.Olsson & 1987 & 21.6  & R.P.Devito & 1981 & 30.3 \\
          & E.L.Hjort & 1994 & 65.0    & J.H.Osborne & 2004 & 107.5,~185.0 \\
\hline
    22-Ti-48 & A.B.Smith & 1998 & 4.5,~5.5,~6.5,~7.55,~8.08,& C.St.Pierre & 1959 & 14.0 \\
             &           &      & ~8.41,~9.06,~9.5,~9.99 &               &       &\\
\hline
    24-Cr-52 & B.Holmqvist& 1969 & 3,4   & W.E.Kinney & 1974 & 4.34,~4.92,~6.44,~8.56 \\
          & A.B.Smith & 1997 & 7.52  & D.Schmidt & 1998 & 7.95,~9.0,~9.8,~10.79,~11.44, \\
          & N.Olsson & 1987 & 21.6  &    &   & 12.01,~12.7,~13.65,~14.1,~14.76\\
\hline
    25-Mn-55 & B.Holmqvist & 1969 & 2.47,~3.0,~3.49,~4.0,~4.56,& Th.Schweitzer & 1978 & 3.4 \\
           &               &      & 6.09,~7.05,~8.05  &    &   &\\
          & A.Takahashi & 1992 & 14.1  &       &  \\
\hline
    26-Fe-56 & V.M.Morozov & 1972 & 1.8   & W.E.Kinney & 1968 & 4.6,~5.0,~5.56,~6.12,~6.53,~7.55 \\
          & P.Boschung & 1971 & 5.05  & Ruan.Xichao & 2009 & 8.17 \\
          & S.Mellema & 1983 & 11.0,~20,~26 & N.Olsson & 1987 & 21.6 \\
          & T.P.Stuart & 1962 & 24.8  & M.Ibaraki & 2002 & 55.0,~65.0,~75.0 \\
\hline
    27-Co-59 & B.Holmqvist & 1969 & 1.46,~2.0,~2.47,~3.0,~3.49,& M.M.Nagadi & 2003 & 9.95,~15.43,~16.88,~18.86 \\
             &             &      & 4.0,~4.56,~6.09,~7.05,~8.05 &            &     & \\
          & L.F.Hansen  & 1985 & 14.6  & N.Olsson  & 1987 & 21.6 \\
          & S.T.Lam & 1985 & 23.0    &       &  \\
\hline
    28-Ni-58 & B.Holmqvist  & 1969 & 3.0     & W.E.Kinney & 1974 & 4.34,~6.44,~7.54,~8.56 \\
          & A.B.Smith & 1992 & 5.5,~6.5,~8.4,~9.5,~9.99 & P.P.Guss & 1985 & 7.9,~9.96,~13.94 \\
          & E.G.Christodoulo & 1999 & 14.0    & A.Takahashi & 1992 & 14.1 \\
          & R.S.Pedroni & 1988 & 16.9  & N.Olsson & 1987 & 21.6 \\
          & Y.Yamanouti & 1979 & 24.0    &       &  \\
\end{tabular}%
\end{ruledtabular}
\end{table*}%

\begin{table*}
\centering \caption{\label{tab2} The $d\sigma/d\Omega$ database for neutron elastic scattering (continued)}
\begin{ruledtabular}
\begin{tabular}{lllllll}
    \toprule
    \textbf{Target} & \textbf{Author(1$^{st}$)} & \textbf{Year} & \textbf{Energy(MeV)} & \textbf{Author(1$^{st}$)} & \textbf{Year} & \textbf{Energy(MeV)} \\
\hline
    29-Cu-63 & P.Guenther & 1986 & 1.6,~2,3,~3.9 & W.E.Kinney & 1974 & 5.5,~7.0,~8.5 \\
          & S.M.El-Kadi & 1982 & 7.96,~9.94,~11.93,~13.92 & J.D.Anderson & 1959 & 14.6 \\
          & B.Ya.Guzhovskiy & 1961 & 15.0    & A.Begum & 1979 & 16.1 \\
          & A.Bratenahl & 1950 & 84.0    & G.L.Salmon & 1960 & 96.0 \\
          & C.P.Van.Zyl & 1956 & 136.0   &       &  \\
\hline
    34-Se-80 & R.M.Musaelyan & 1987 & 0.34  & E.S.Konobeevskij & 1984 & 1.19 \\
          & I.A.Korzh & 1983 & 1.5,~2.0,~2.5,~3,5 & G.V.Gorlov & 1964 & 4 \\
          & R.G.Kurup & 1984 & 8.0,~10.0  &       &  \\
\hline
    38-Sr-88 & S.A.Cox  & 1972 & 0.886 & M.Walt & 1954 & 1 \\
          & D.W.Kent & 1962 & 3.66  & V.I.Popov & 1971 & 4.37 \\
          & D.E.Bainum & 1978 & 11    &       &  \\
\hline
    39-Y-89 & R.D.Lawson & 1986 & 4.5,~5.0,~5.5,~5.9,~6.5,~7.14, & F.G.Perey & 1970 & 7.6,~8.56 \\
            &            &      & 7.5,~8.03,~8.4,~9.06,~9.5,~9.99 &  &  \\
          & G.M.Honore & 1986 & 7.96,~9.95,~11.94,~13.93 & S.Mellema & 1987 & 11.0 \\
          & N.Olsson & 1987 & 21.6  &       &  \\
\hline
    40-Zr-90 & P.Guenther & 1975 & 2.0,~2.2,~2.6,~3,4 & R.W.Stooksberry & 1976 & 2.11 \\
          & S.Chiba & 1992 & 4.5,~5.0,~5.5,~5.9,~6.5,& Y.Wang & 1990 & 10.0,~24.0 \\
          &         &      & 8.03,~9.06,~9.99 \\
          & D.E.Bainum & 1978 & 11.0    & M.Ibaraki & 2002 & 55.0,~65.0,~75.0 \\
\hline
    41-Nb-93 & A.B.Smith & 1985 & 4.5,~5.0,~5.5,~5.9,~6.5,~7.14& R.S.Pedroni & 1991 & 7.95,~9.94,~11.93,~13.92,~16.91 \\
             &           &      & 7.5,~8.03,~8.4,~9.06     &  & \\
          & J.C.Ferrer & 1977 & 11.0    & E.G.Christodoulo & 1999 & 14.0 \\
          & R.Finlay & 1991 & 20.0    &       &  \\
\hline
    42-Mo-98 & P.Lambropoulos  & 1973 & 1.5 & A.B.Smith & 1975 & 2.0,~3.0,~4.0 \\
          & J.Rapaport & 1979 & 7.0,~9.0,~11.0,~16.0,~20.0,~26.0&       &  \\
\hline
    45-Rh-103 & A.B.Smith & 1994 & 4.51,~5.0,~5.9,~6.5,~7.5,\\
              &           &      & 8.03,~8.4,~9.06,~9.5,~10.0 \\
\hline
    49-In-115 & S.A.Cox & 1972 & 0.87  & B.Holmqvist & 1969 & 3.0,~4.0,~7.05,~8.05 \\
          & A.B.Smith & 1984 & 3.05,~3.75 & R.L.Becker & 1966 & 3.2 \\
          & S.Chiba & 1990 & 4.5,~5.0,~5.9,~7.14,~8.03, & J.C.Ferrer & 1977 & 11.0 \\
          &         &      & 9.06,~9.99 & & \\
          & J.O.Elliot & 1956 & 14.0    & L.F.Hansen & 1985 & 14.6 \\
\hline
    50-Sn-120 & S.Tanaka & 1972 & 1.52,~2.05,~2.57,~3.08 & C.Budtz-Jorgense & 1984 & 3.0,~3.2,~3.4,~3.6,~3.8,~4.0 \\
          & R.M.Wilenzick & 1965 & 6.04  & P.P.Guss & 1989 & 9.94,~13.92,~16.91 \\
          & J.Rapaport & 1980 & 11.0    & T.P.Stuart & 1962 & 24.0 \\
          & E.L.Hjort & 1994 & 65.0    &       &  \\
\hline
    79-Au-197 & R.B.Day & 1965 & 0.5,~2.5 & F.T.Kuchnir & 1968 & 0.6,~1.6 \\
          & S.A.Cox & 1972 & 0.878,~2.0 & A.B.Smith & 2005 & 4.51,~5.51,~6.51,~7.51,~8.41,~9.99 \\
          & S.C.Buccino & 1966 & 5     & M.A.Etemad & 1973 & 7 \\
          & B.Holmqvist & 1971 & 8.05  & L.F.Hansen & 1985 & 14.6 \\
\hline
    82-Pb-208 & V.M.Morozov & 1972 & 1.8   & J.R.M.Annand & 1985 & 4.0,~5.0,~6,7 \\
          & D.Schmidt & 1996 & 7.93,~8.98,~9.87,~10.96,~11.92,& W.E.Kinney & 1974 & 8.5 \\
          &           &      & 13.12,~14.23 \\
          & J.Rapaport & 1978 & 11.0,~26.0 & A.Takahashi & 1987 & 14.1 \\
          & R.W.Finlay & 1984 & 20.0,~22.0,~24.0 & R.P.Devito & 1980 & 30.3,~40.0 \\
          & M.Ibaraki & 2002 & 55,~65.0,~75.0 & A.Bratenahl & 1950 & 84.0 \\
          & J.H.Osborne & 2004 & 85.0,~95.0,~107.0,~127.5,~155.0,& A.Oehrn & 2008 & 96.0 \\
          &             &      & 185.0,~225.0 \\
          & C.P.Van.Zyl & 1956 & 136.0   &       &  \\
\hline
    83-Bi-209 & N.Olsson & 1982 & 1.48,~1.97,~2.23,~3.05 & J.R.M.Annand & 1985 & 4.0,~5.0,~5.5,~6.5,~7.0 \\
          & R.K.Das & 1990 & 7.5,~8.0,~9.0,~10.0,~11.0,& N.Olsson & 1987 & 21.6 \\
          &         &      & 12.0,~20.0,~24.0 \\
\end{tabular}%
\end{ruledtabular}
\end{table*}%

\subsubsection{The neutron total cross section}

\begin{figure}[htbp]
\centerline{\includegraphics[width = 3.5in]{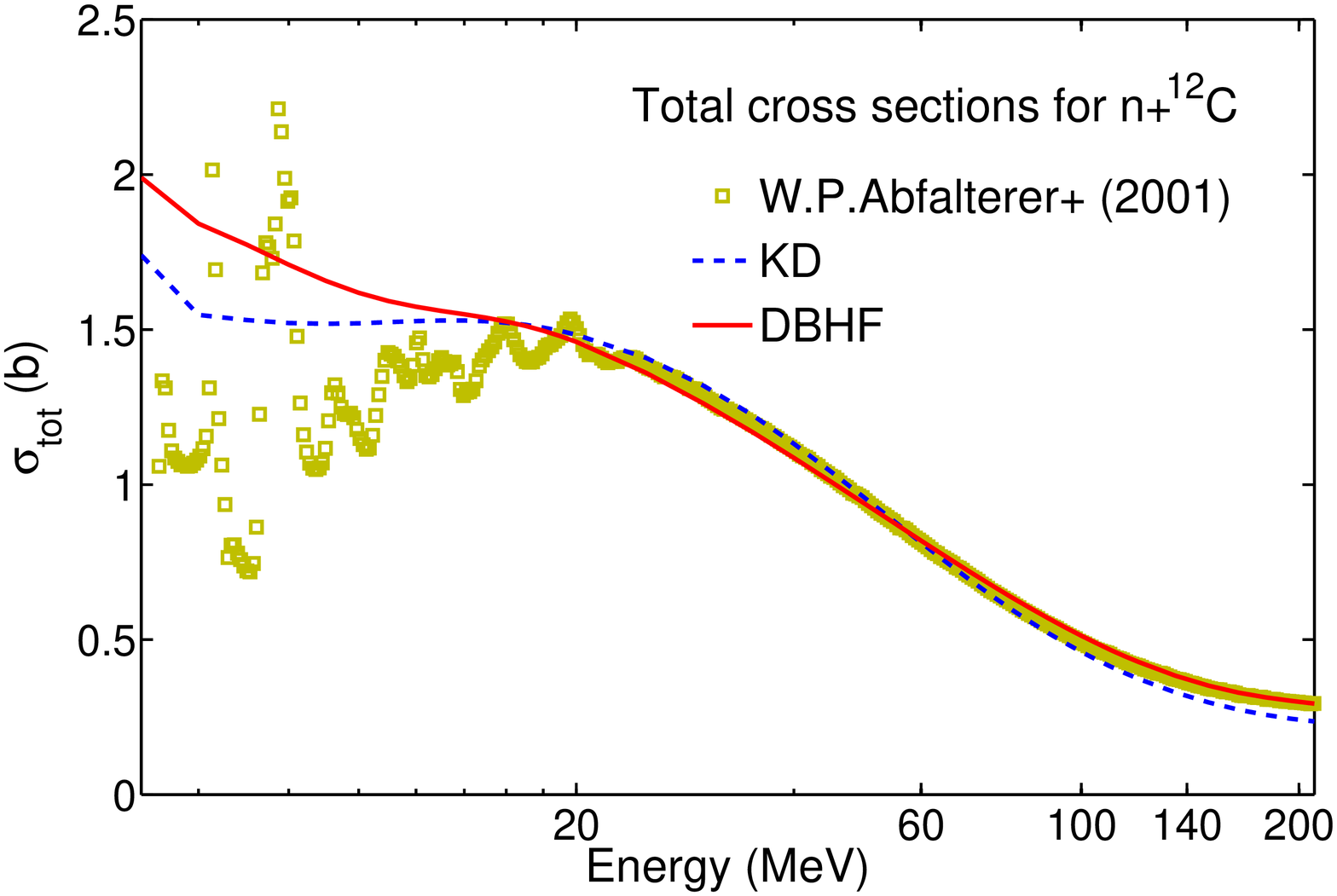}}
\caption{(color online) Comparison of predicted neutron total cross section (solid line) and experimental data (point) and KD calculation (dashed line) for $n$ +$^{12}$C.
The experimental data are measured for natural carbon.}
\label{fig3}
\end{figure}

\begin{figure}[htbp]
\centerline{\includegraphics[width = 3.5in]{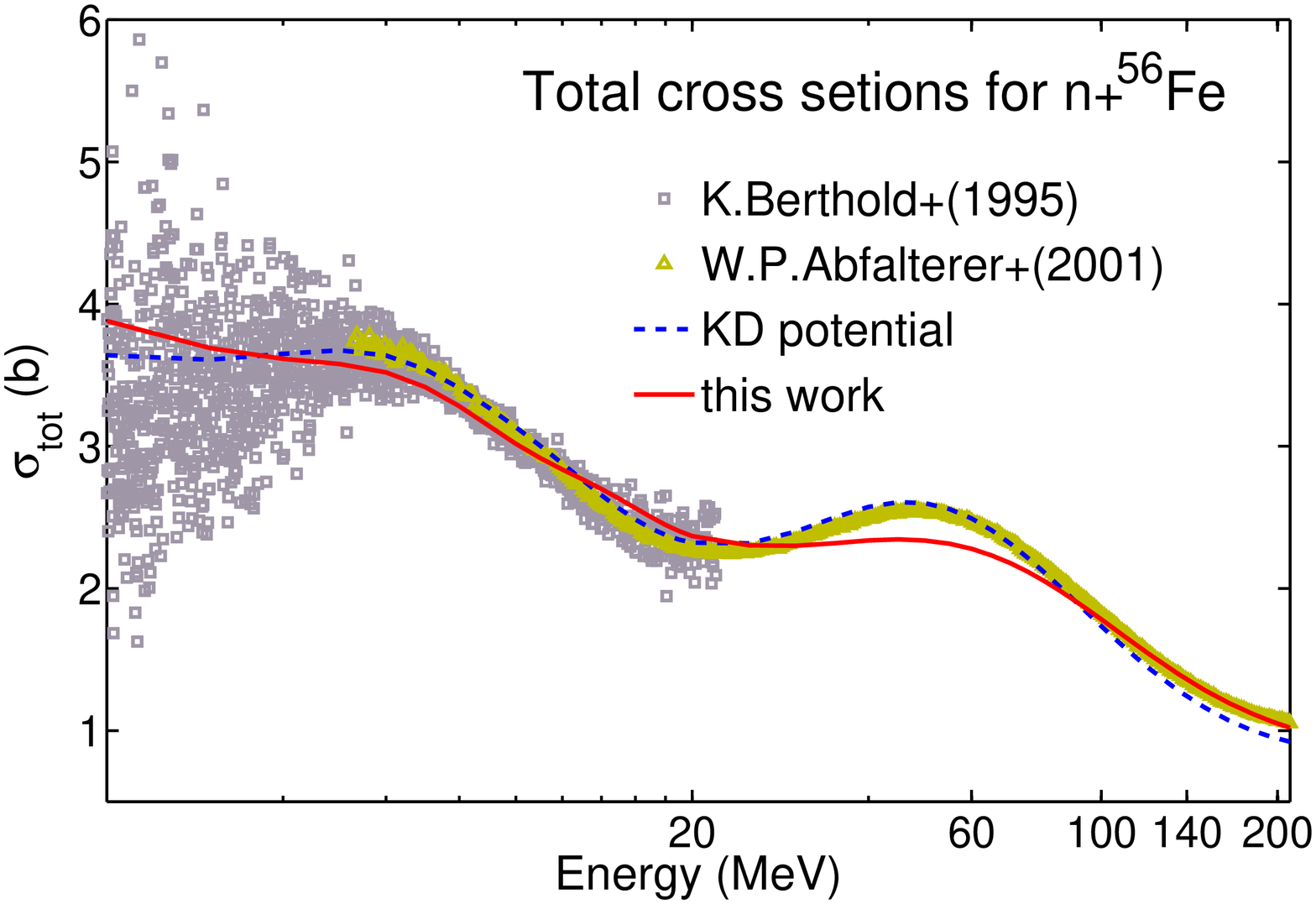}}
\caption{(color online) Comparison of predicted neutron total cross section (solid line) and experimental data (point) and KD calculation (dashed line) for $n$ +$^{56}$Fe.
The experimental data are measured for natural iron.}
\label{fig4}
\end{figure}

\begin{figure}[htbp]
\centerline{\includegraphics[width = 3.5in]{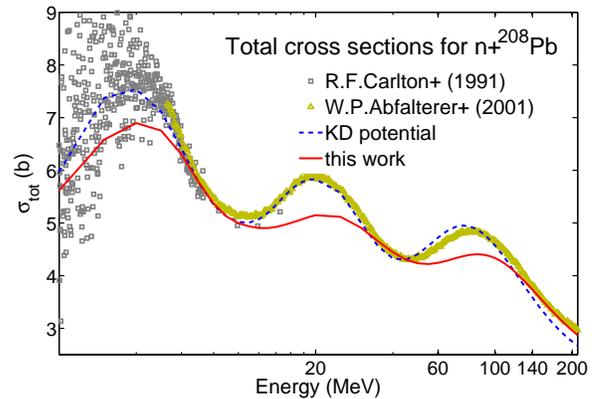}}
\caption{(color online) Comparison of predicted neutron total cross section (solid line) and experimental data (point) and KD calculation (dashed line) for $n$ +$^{208}$Pb.
The experimental data are measured for natural lead.}
\label{fig5}
\end{figure}

The calculated neutron total cross sections of $^{12}$C, $^{56}$Fe
and $^{208}$Pb are compared with the experimental data and the
results of KD potential in Figs. \ref{fig3}, \ref{fig4} and
\ref{fig5}. Within the scope of application (E$_n$ $>$ 30 MeV), a
satisfactory prediction is obtained for light nucleus $^{12}$C.
Because more Ramsauer-like structures appear for the heavy nuclei,
the data quality of prediction decrease with increasing mass number.
The cross sections are underestimated in this work. The most
deviation between experimental data and calculation reaches 10\% for
$^{208}$Pb.

\subsubsection{The elastic scattering angular distribution}

\begin{figure*}[htbp]
\centerline{\includegraphics[width = 5in]{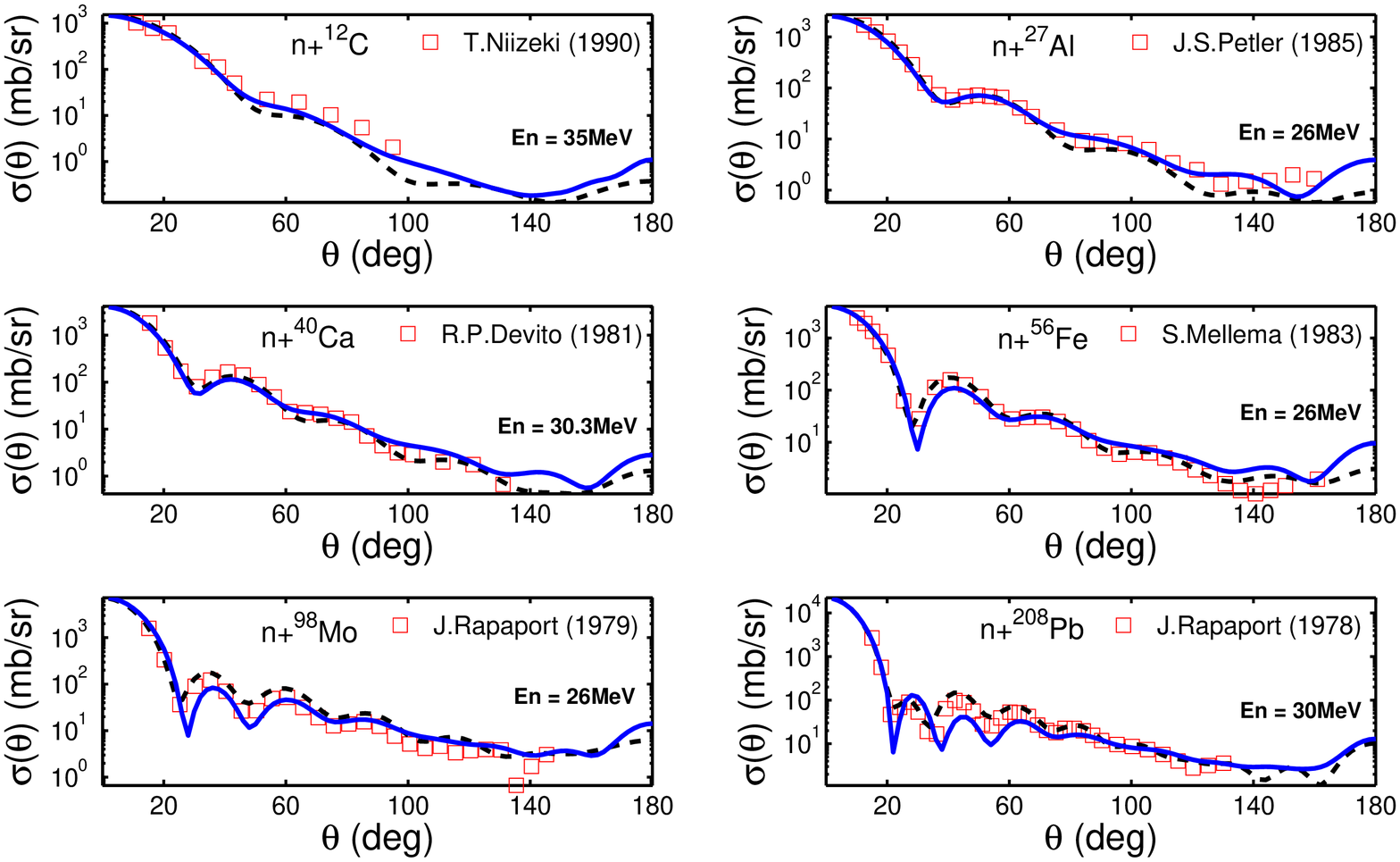}}
\caption{(color online) Comparisons of angular distributions for $n$
+ $^{12}$C,$^{27}$Al,$^{40}$Ca,$^{56}$Fe,$^{98}$Mo and $^{208}$Pb at
incident neutron energy around 30MeV. The dashed line indicates the
results from KD potential and the solid line denotes the present
prediction.} \label{fig6}
\end{figure*}

\begin{figure*}[htbp]
\centerline{\includegraphics[width = 5in]{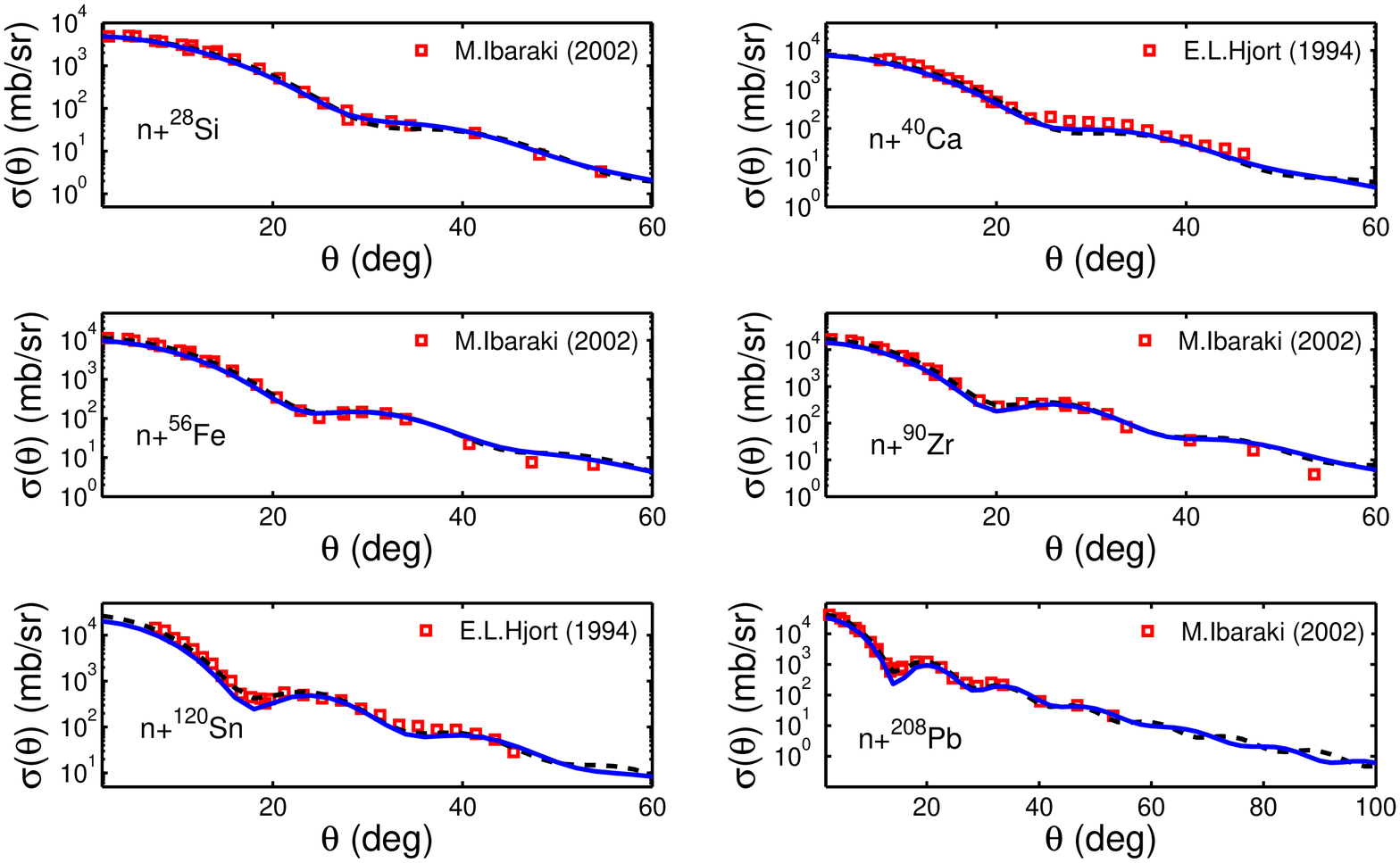}}
\caption{(color online) Comparisons of angular distributions for $n$
+ $^{28}$Si,$^{40}$Ca,$^{56}$Fe,$^{90}$Zr,$^{120}$Sn and $^{208}$Pb
at incident neutron energy around 65MeV. The dashed line indicates
the results from KD potential and the solid line denotes the present
prediction.} \label{fig7}
\end{figure*}

As abundant experimental data are existed, we show more concern on
$d\sigma/d\Omega$ in this MOP study. Overall, the predicted results
are satisfactory even below the energy scope of application of MOP.
As the examples, the calculated $d\sigma/d\Omega$ around incident
neutron at 30MeV and 65MeV are plotted individually in Figs. \ref{fig6} and \ref{fig7}, where the present predictions coincide
with experimental data and the KD results very well. More details
for diversified nuclei are exhibited in the following contents.

\begin{figure}[htbp]
\centerline{\includegraphics[width = 4.0in]{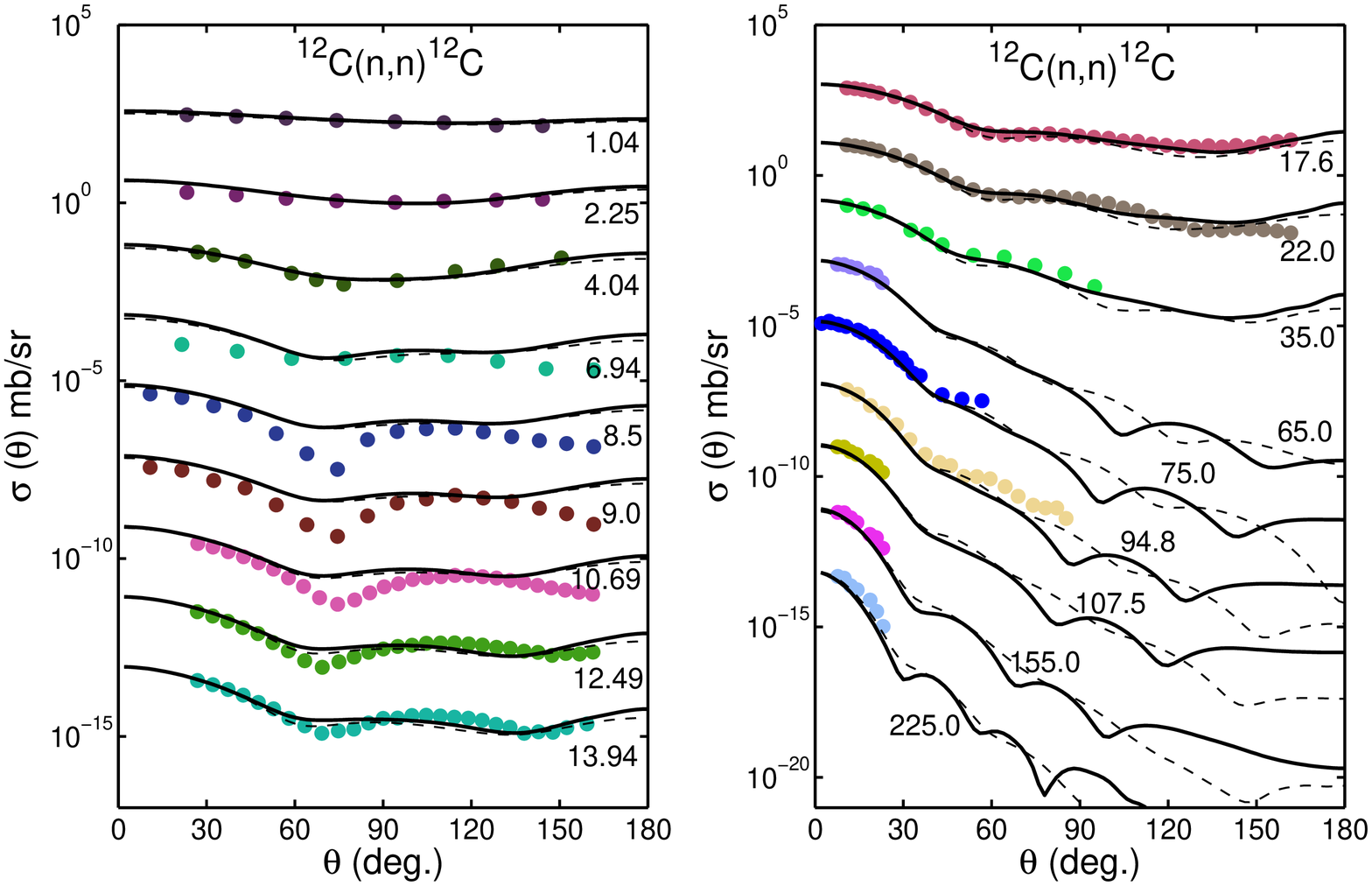}} \caption{
(color online) Comparison of predicted $d\sigma/d\Omega$ (solid
line) and experimental data (point) and KD calculation (dashed line)
for $n$ + $^{12}$C.}
\label{fig8}
\end{figure}

Note that for neutron elastic differential cross sections, as,
e.g., in Fig. \ref{fig8}, the incident laboratory energies are in
MeV. The curves and data points at the top are true values, while
the others are offset by factors of 0.01, 0.0001, etc.

 \textbf{Targets $^{12}$C - $^{40}$Ca:} 9 target nuclei including $^{12}$C, $^{14}$N, $^{16}$O,
$^{23}$Na, $^{24}$Mg, $^{27}$Al, $^{28}$Si, $^{32}$S, and $^{40}$Ca
are examined in this mass region. Only $^{40}$Ca was considered in
the optimization procedure as the example for an isospin symmetric
nucleus while the results obtained for the other nuclei can be
considered as predictions  of our model. The resulting $\chi^{2}/N$
are tabulated in Table \ref{tab3}. The results for all nuclei are
smaller than 1.0, except for the target nucleus $^{12}$C. It is
remarkable that the results obtained for the phenomenological KD
potential exhibit the same trends as can be observed in the
microscopic optical potential. For some target nuclei like e.g.
$^{27}$Al both models yield a very small value for $\chi^2/N$ while
both models yield a rather poor result for other nuclei, like e.g.
$^{12}$C. The scattering on such nuclei is very much influenced by
the existence of specific surface excitation modes, which cannot be
described in terms of a global optical model (see discussion above).
It is worth mentioning that the value for $\chi^2/N$ for $^{40}$Ca,
which has been included in the fit procedure, is comparable to the
corresponding value for the other nuclei, which have not been
considered in the optimization procedure.

\begin{figure}[htbp]
\centerline{\includegraphics[width = 4.0in]{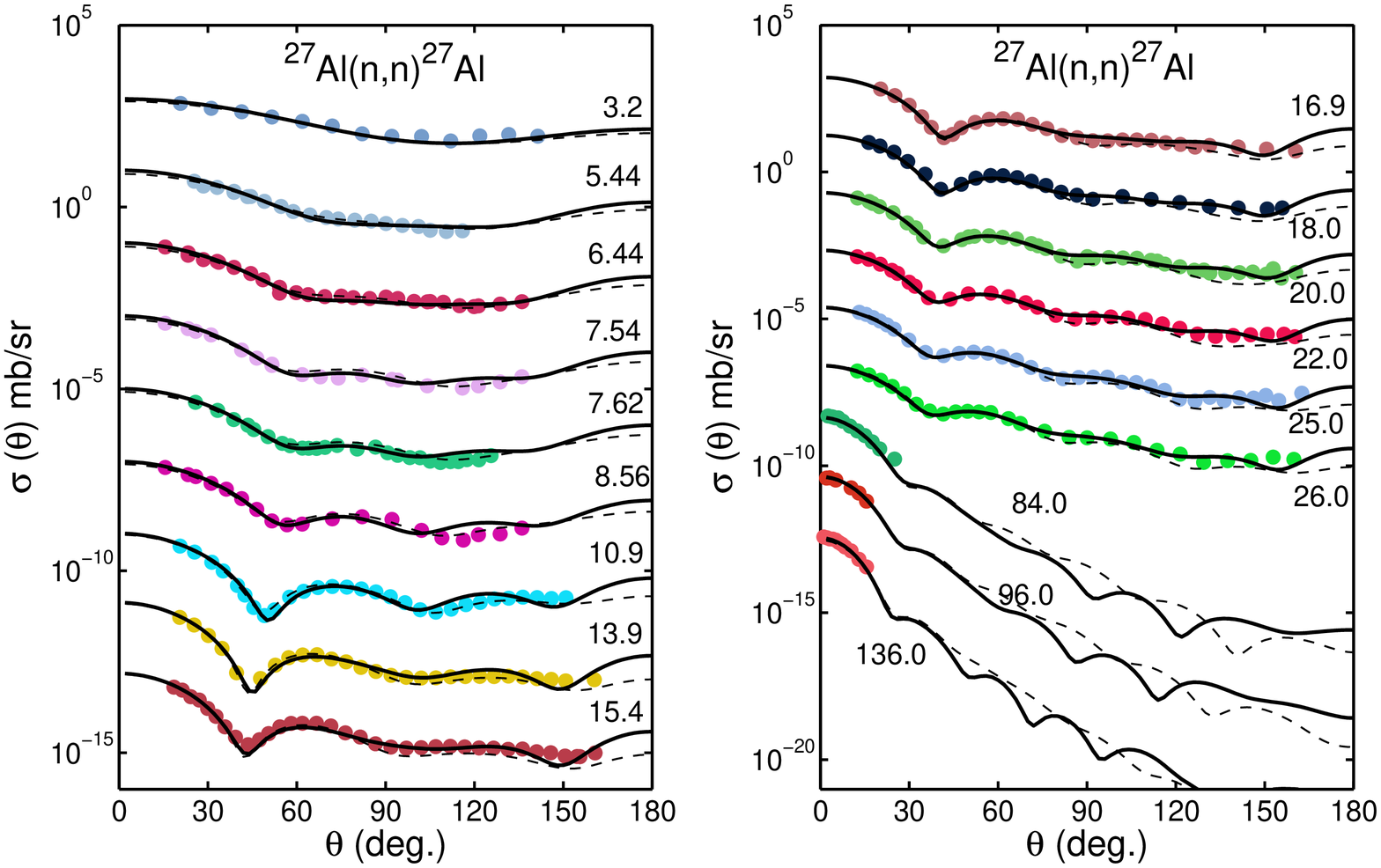}} \caption{
(color online) Comparison of predicted $d\sigma/d\Omega$ (solid
line) and experimental data (point) and KD calculation (dashed line)
for $n$+ $^{27}$Al.}
\label{fig9}
\end{figure}

\begin{figure}[htbp]
\centerline{\includegraphics[width = 4.0in]{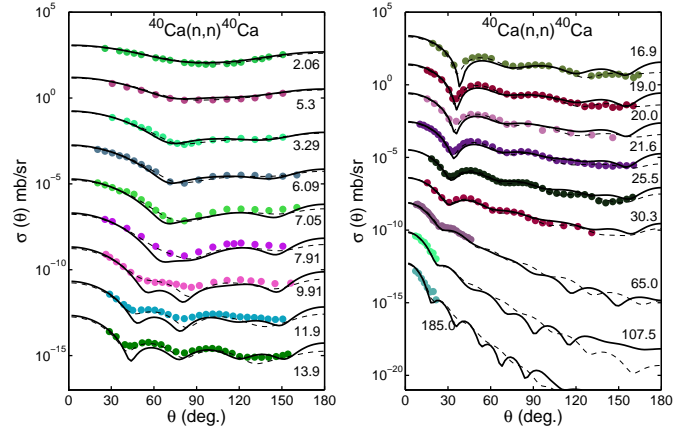}} \caption{
(color online) Comparison of predicted $d\sigma/d\Omega$ (solid
line) and experimental data (point) and KD calculation (dashed line)
for $n$+ $^{40}$Ca.}
\label{fig10}
\end{figure}

The visual comparisons of the present predictions for $^{12}$C,
$^{27}$Al and $^{40}$Ca with the experimental data, as well as those
with KD potential are shown in Figs. \ref{fig8}, \ref{fig9} and
\ref{fig10}. An excellent agreement with experimental data for $n$ +
$^{27}$Al is observed in Fig. \ref{fig9}, as already indicated in
the corresponding value for $\chi^{2}/N$  in Table \ref{tab3}. From
the results displayed in Fig. \ref{fig8} one can see that main
contributions to the large value of $\chi^2/N$ for $^{12}$C originate
from the deviations between measurements and theoretical results at
the energies around E$_n$ = 7 MeV to 13 MeV. The results for $^{40}$Ca by
MOP is good except for a slight underestimation at energies 10 MeV - 20 MeV
around the angles between 30$^o$ to 60$^o$, while the
phenomenological results describe the data in this region in a very
reasonable way.

\begin{table}
\caption{\label{tab3} The $\chi^{2}/N$ of $d\sigma/d\Omega$ for $n+^{12}$C - $^{40}$Ca reactions}
\begin{ruledtabular}
\begin{tabular}{cccc}
 Nuclide &  N of data points &      MOP  & KD\\
\hline
 $^{12}$C & 293 & 3.35 & 2.43\\
\hline
 $^{14}$N & 336 & 0.21& 0.22\\
\hline
 $^{16}$O & 309 & 0.91& 0.66\\
\hline
 $^{23}$Na & 221 & 0.31& 0.22\\
\hline
 $^{24}$Mg & 270 & 0.56& 0.19\\
\hline
 $^{27}$Al & 426 & 0.068& 0.069\\
\hline
 $^{28}$Si & 391 & 0.24& 0.15\\
\hline
 $^{32}$S & 388 & 0.22& 0.07\\
\hline
 $^{40}$Ca & 399 & 0.22& 0.075\\
\end{tabular}
\end{ruledtabular}
\end{table}

\begin{figure}[htbp]
\centerline{\includegraphics[width = 4.0in]{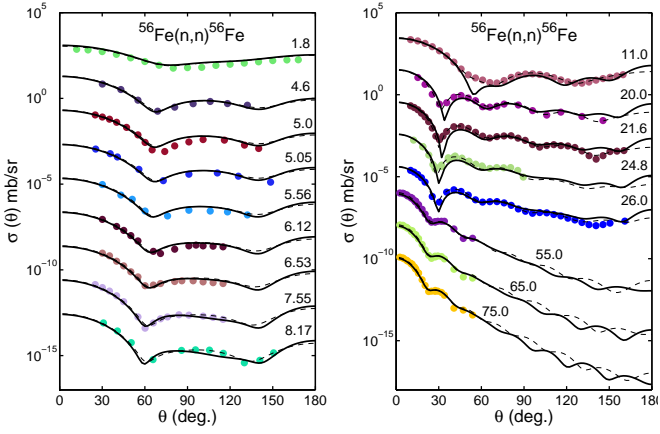}} \caption{
(color online) Comparison of predicted $d\sigma/d\Omega$ (solid
line) and experimental data (point) and KD calculation (dashed line)
for $n$+ $^{56}$Fe.} \label{fig11}
\end{figure}

\textbf{Targets $^{48}$Ti-$^{63}$Cu:} We compare results for 5
nuclei in this mass region, which are important components of
structure materials:  $^{48}$Ti, $^{52}$Cr, $^{56}$Fe, $^{58}$Ni,
and $^{63}$Cu. The values of $\chi^{2}/N$ are suspended around
0.11-0.17 except for a slightly larger value of 0.23 for $^{56}$Fe
(see Table \ref{tab4}). As an example we show our prediction for
$^{56}$Fe in Fig. \ref{fig11} and compare it with the experimental
data and the results of corresponding calculations using the
phenomenological KD model. Over all our results show a fairly good
agreement with the experimental data. The largest discrepancies
occur for incident energies around 10 MeV to 20 MeV in a region of
scattering angles between 30$^o$ - 90$^o$. In fact, this deviation appears
throughout this mass region.

\begin{table}
\caption{\label{tab4} The $\chi^{2}/N$ of $d\sigma/d\Omega$ for
$n+^{48}$Ti - $^{63}$Cu reactions}
\begin{ruledtabular}
\begin{tabular}{cccc}
 Nuclide &  Point num. of exp. &  MOP  & KD\\
\hline
 $^{48}$Ti & 378 & 0.13 & 0.05\\
\hline
 $^{52}$Cr & 562 & 0.16& 0.03\\
\hline
 $^{56}$Fe & 333 & 0.23& 0.09\\
\hline
 $^{58}$Ni & 701 & 0.17& 0.11\\
\hline
 $^{63}$Cu & 282 & 0.11& 0.06\\
\end{tabular}
\end{ruledtabular}
\end{table}

\begin{figure}[htbp]
\centerline{\includegraphics[width = 4.0in]{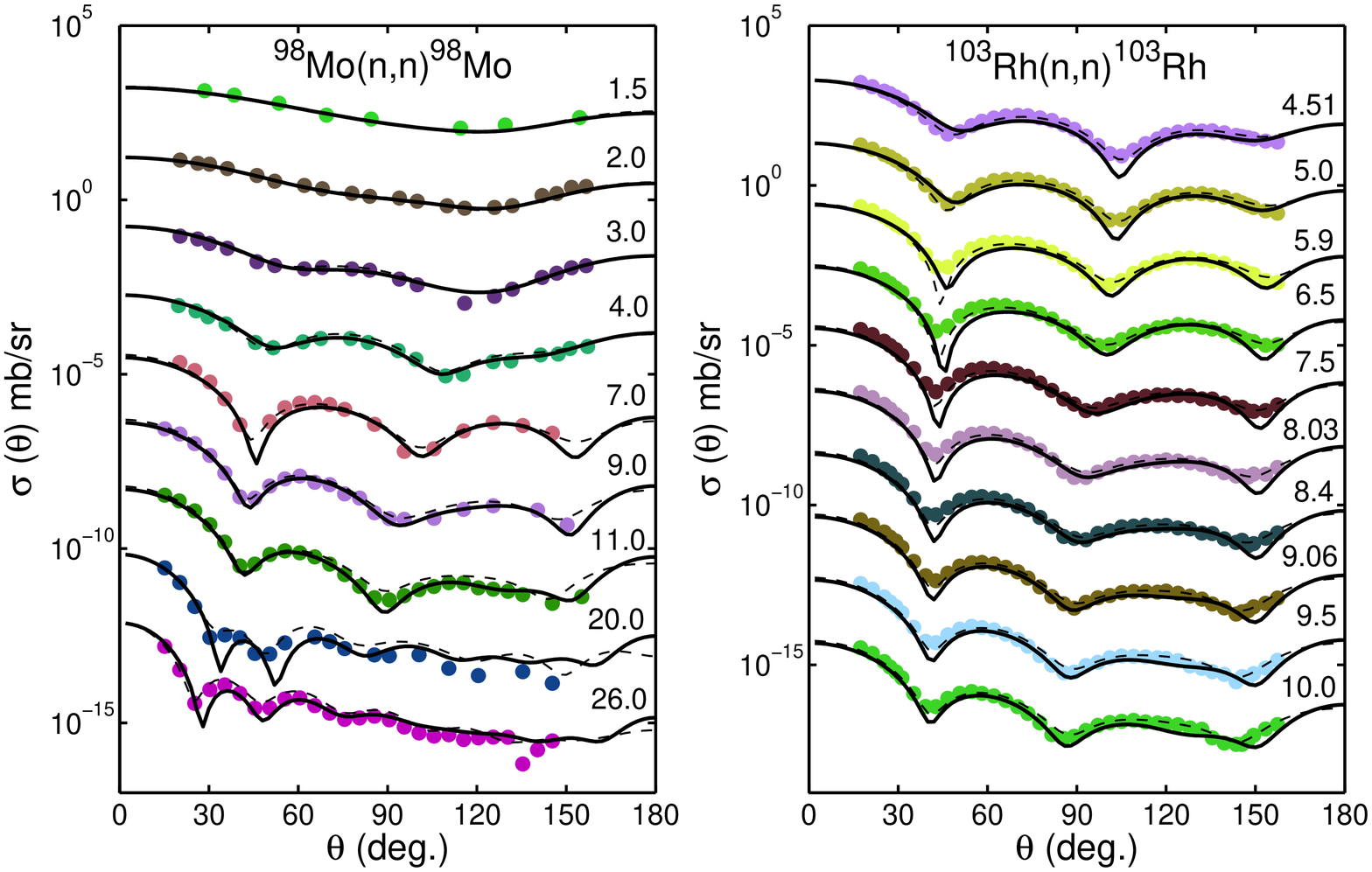}} \caption{
(color online) Comparison of predicted $d\sigma/d\Omega$ (solid line) and experimental data (point)
and KD calculation (dashed line) for $n$ + $^{98}$Mo and $^{103}$Rh.}
\label{fig12}
\end{figure}

\begin{figure}[htbp]
\centerline{\includegraphics[width = 4.0in]{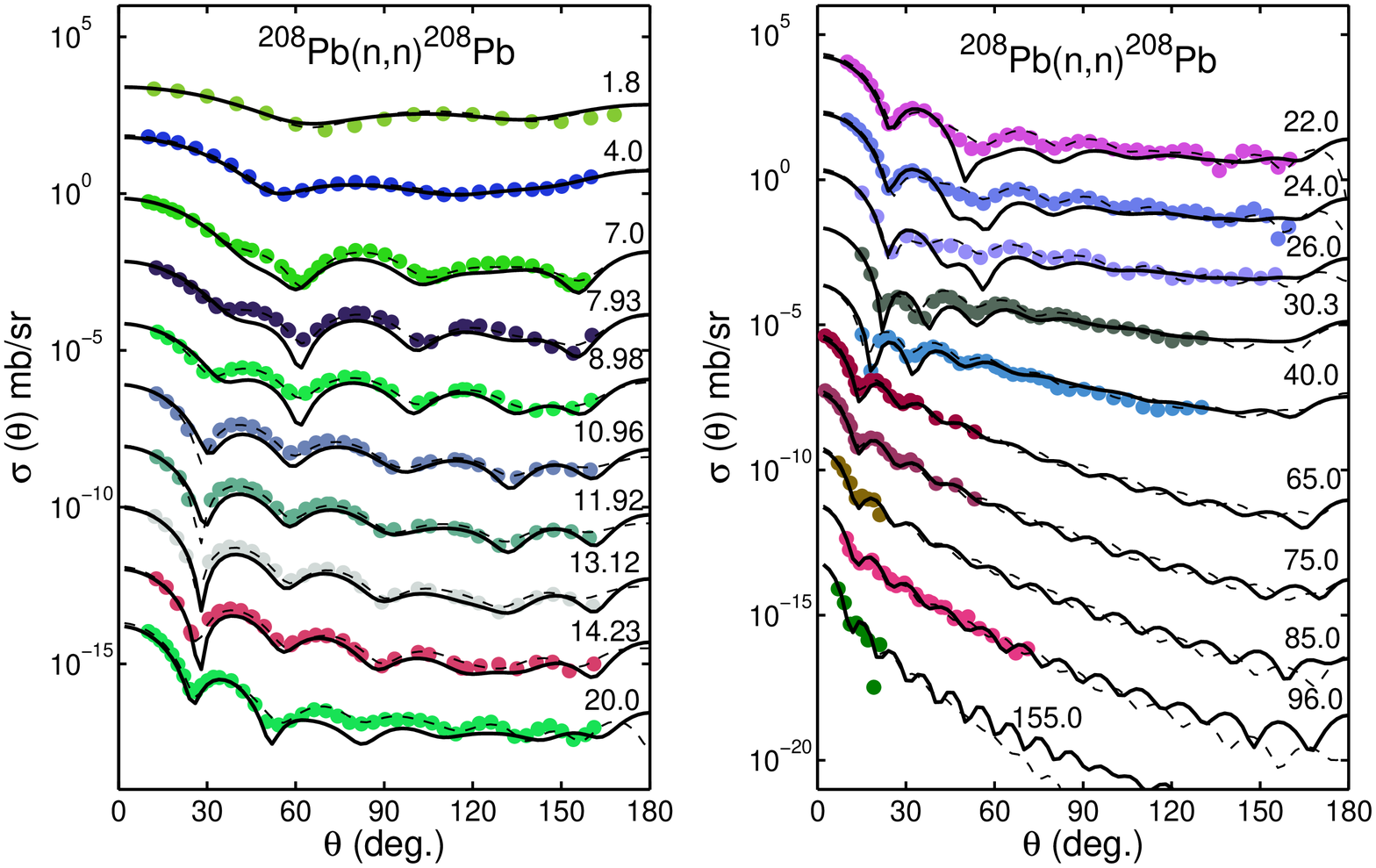}} \caption{
(color online) Comparison of predicted $d\sigma/d\Omega$ (solid line) and experimental data (point)
and KD calculation (dashed line) for $n$ + $^{208}$Pb.}
\label{fig13}
\end{figure}

 \textbf{Targets $^{80}$Se-$^{209}$Bi:} 13 nuclei including $^{80}$Se, $^{88}$Sr, $^{89}$Y, $^{90}$Zr,
 $^{93}$Nb, $^{98}$Mo, $^{103}$Rh, $^{115}$In, $^{120}$Sn, $^{140}$Ce,
$^{197}$Au, $^{208}$Pb and $^{209}$Bi are utilized to test the
performance of this MOP. Good agreement is obtained generally, which
could be perceived through the criteria $\chi^{2}/N$ in Table
\ref{tab5} and Figs. \ref{fig12} and \ref{fig13} for $^{98}$Mo,
$^{103}$Rh and $^{208}$Pb. It is noticed that the deviation in the
minimum of the angular distribution at scattering angels 30$^{o}$ - 60$^{o}$
around E$_{n}$ = 20 MeV, which has been discussed above for
$^{48}$Ti-$^{63}$Cu also shows up for these nuclei. The results near
the incident energy 20 MeV - 30 MeV generally exhibit the
underestimation around 50$^o$, which is illustrated also by
$d\sigma/d\Omega$ for $^{208}$Pb in Fig. \ref{fig13}. Apart from
these defects above, all the $d\sigma/d\Omega$ for other heavy
target nuclei are reproduced in a very nice way.

\begin{table}
\caption{\label{tab5} The $\chi^{2}/N$ of $d\sigma/d\Omega$ for
$n+^{80}$Se-$^{209}$Bi reactions}
\begin{ruledtabular}
\begin{tabular}{cccc}
 Nuclide &  Point num. of exp. &  MOP  & KD\\
\hline
 $^{80}$Se & 152 & 0.17 & 0.11\\
\hline
 $^{88}$Sr & 81 & 0.09 & 0.03\\
\hline
 $^{89}$Y & 620 & 0.19 & 0.05\\
\hline
 $^{90}$Zr & 1110 & 0.14& 0.05\\
\hline
 $^{93}$Nb & 629 & 0.13& 0.03\\
\hline
 $^{98}$Mo & 180 & 0.30 & 0.36\\
\hline
 $^{103}$Rh & 400 & 0.12 & 0.06\\
\hline
 $^{115}$In & 744 & 0.10 & 0.05\\
\hline
 $^{120}$Sn & 357 & 0.08& 0.03\\
\hline
 $^{140}$Ce & 105 & 0.19 & 0.05\\
\hline
 $^{197}$Au & 390 & 0.22 & 0.10\\
\hline
 $^{208}$Pb & 885 & 2.25 & 1.80\\
\hline
 $^{209}$Bi & 767 & 0.27 &  0.06\\
\end{tabular}
\end{ruledtabular}
\end{table}

\subsubsection{The analyzing power}

\begin{figure*}[htbp]
\centerline{\includegraphics[width =
5in]{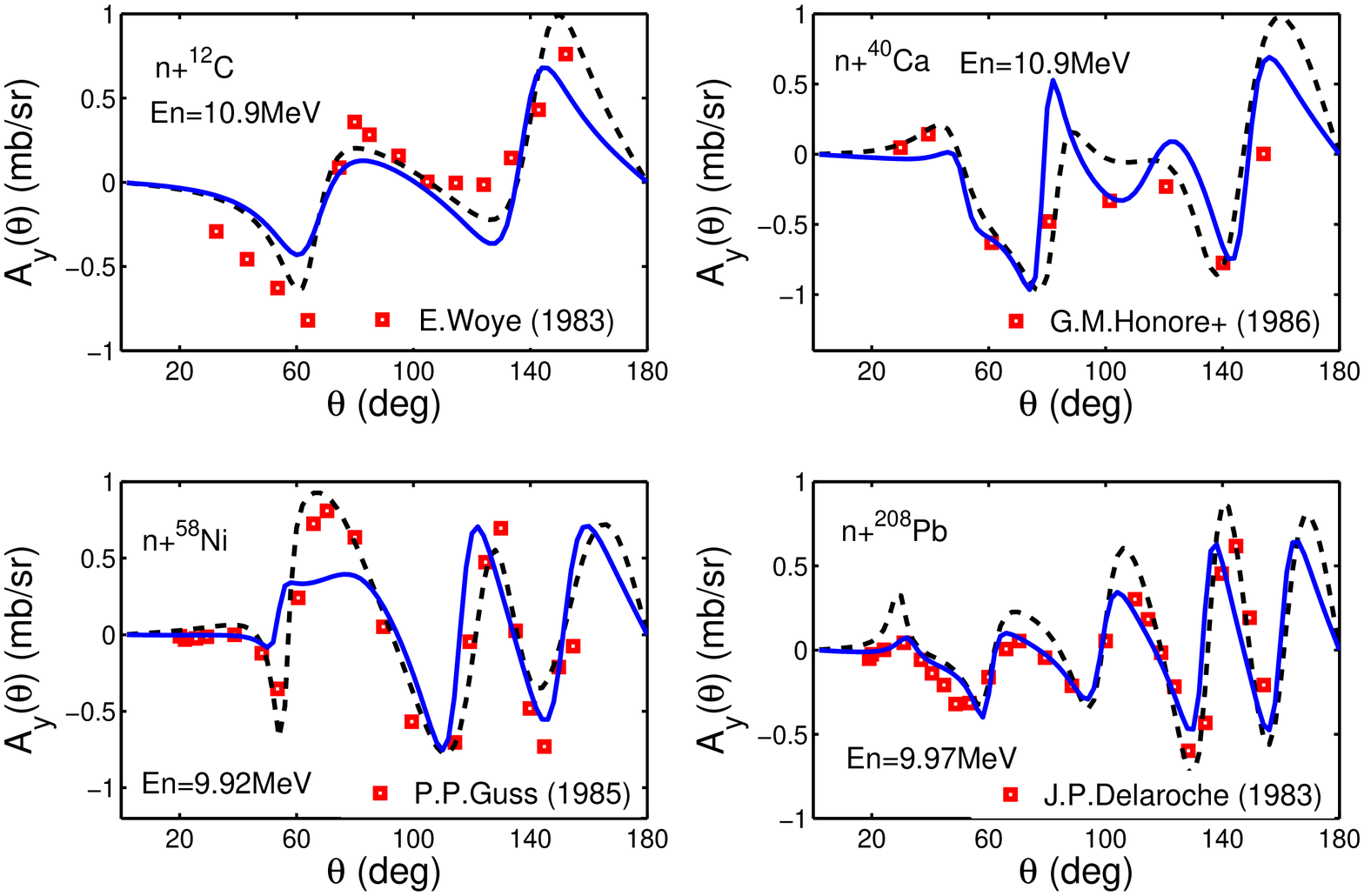}} \caption{(color online)
Comparisons of analyzing power for $n$ + $^{12}$C, $^{40}$Ca,
$^{58}$Ni, and $^{208}$Pb at incident neutron energy around 10MeV.
The dashed line indicates the results from KD potential and the
solid line denotes the present work.}
\label{fig14}
\end{figure*}

As mentioned above, it is the important feature of the relativistic
description that the spin-orbit term can be naturally involved in
the scheme without any additional parameter, which is beneficial to
derive the spin-orbit observables A$_y$($\theta$) and
Q$_y$($\theta$). The A$_y$($\theta$) at incident energies around 10
MeV are selected to show the ability of predictions for $^{12}$C,
$^{40}$Ca, $^{58}$Ni and $^{208}$Pb in Fig. \ref{fig14}, and good
agreements with the experimental data for all nuclei are obtained.

\subsection{Results for proton scattering}

About 150 elastic scattering angular distributions, 65 analyzing
powers, and reaction cross sections of 10 commonly targets have been
included in our systematic comparison. The experimental database of
$d\sigma/d\Omega$ is summarized in Table \ref{tab6} and depicted by
the first author there. As for other quantities, the experimental
data for plotting are introduced in the figures.

We compare the present calculations with the experimental data and
KD results. As a whole, the various proton scattering observables
are predicted satisfactorily using MOP just like its performance in
neutron scattering reactions, the results are discussed in the
following subsections.

\begin{table*}[htbp]
  \centering
  \caption{The $d\sigma/d\Omega$ database for proton elastic scattering}
    \begin{ruledtabular}
    \begin{tabular}{lllllll}
    \textbf{Target} & \textbf{Author(1$^{st}$)} & \textbf{Year} & \textbf{Energy(MeV)} & \textbf{Author(1$^{st}$)} & \textbf{Year} & \textbf{Energy(MeV)} \\
\hline
    6-C-12 & S.Mazzoni & 1998 & 2.5   & V.M.Lebedev & 2006 & 7.5 \\
          & An-Zhu & 2003 & 22.0    & M.Harada & 1999 & 26.0 \\
          & M.Ieiri & 1987 & 29.7,~34.5,~44.7 & V.I.Grancev & 1983 & 48.5 \\
          & A.A.Rush & 1971 & 50.0    & M.Ieiri & 1987 & 54.4,~64.9,~74.8,~83.8 \\
          & H.O.Meyer & 1983 & 122.0,~160.0,~200.0,~250 & V.M.Hannen & 2003 & 150.0 \\
\hline
    13-Al-27 & M.Chiari & 2001 & 0.783,~1,2,~3.01 & I.E.Dayton & 1956 & 17.0 \\
          & G.M.Crawley & 1968 & 17.5  & R.Dittman & 1969 & 28.0 \\
          & C.B.Fulmer & 1969 & 61.4  & G.Gerstein & 1957 & 92.9,~95.7 \\
          & A.E.Taylor & 1961 & 142.0   & V.Comparat & 1974 & 156.0 \\
          & A.Johansson & 1960 & 160.0,~177.0,183.0 & S.Dahlgren & 1967 & 185.0 \\
\hline
    14-Si-28 & E.Fabrici & 1980 & 14.26,~17.24,~20.17,~30.5,~40.21 & M.Nakamura & 1983 & 45.0,~50.0,~55.0,~60.0 \\
          & S.Kato & 1985 & 65.0    & C.Olmer & 1984 & 80.0,~100.0,~135.0,~179.0 \\
          & O.Sundberg & 1967 & 185.0   & K.H.Hicks & 1988 & 200.0,~250.0 \\
\hline
    20-Ca-40 & J.F.Dicello & 1971 & 10.4,~14.5,~17.6,~20.6 & R.H.Mccamis & 1986 & 25.0,~30.0,~35.0,~40.0,~45.0,~48.0 \\
          & K.Yagi & 1964 & 55.0    & H.Sakaguchi & 1982 & 65.0 \\
          & P.Schwandt & 1982 & 80.0,~135.0,~160.0 & C.Rolland & 1966 & 152.0 \\
          & A.Johansson & 1961 & 182.0   & H.Seifert & 1993 & 201.0 \\
\hline
    26-Fe-56 & N.Boukharouba & 1992 & 4.08,~5.02,~6.56,~7.74 & K.Kikuchi & 1959 & 7.4,~14.1 \\
          & J.Benveniste & 1964 & 10.9  & R.Varner & 1986 & 16.0 \\
          & I.E.Dayton & 1956 & 17.0    & P.Kossanyi-Demay & 1967 & 18.6 \\
          & S.F.Eccles & 1966 & 19.1  & B.W.Ridley & 1964 & 30.3 \\
          & M.K.Brussel & 1959 & 39.8  & F.E.Bertrand & 1969 & 61.5 \\
          & H.Sakaguchi & 1982 & 65.0    & D.J.Steinberg & 1964 & 146 \\
          & V.Comparat & 1974 & 156.0   & A.Johansson & 1961 & 176 \\
\hline
    28-Ni-58 & L.L.Lee Jr & 1964 & 7.0,~8.0,~9.0,~10.0,~11.0,~12.0 & S.Kobayashi & 1960 & 14.4,~15.4 \\
          & R.Varner & 1986 & 16.0    & S.F.Eccles & 1966 & 18.6 \\
          & J.R.Tesmer & 1972 & 20.0    & E.Fabrici & 1980 & 35.2 \\
          & L.N.Blumberg & 1966 & 40.0    & C.B.Fulmer & 1969 & 61.4 \\
          & H.Sakaguchi & 1982 & 65.0    & A.Ingemarsson & 1979 & 178.0 \\
          & H.Sakaguchi & 1998 & 192.0   & H.Takeda & 2003 & 250.0 \\
\hline
    40-Zr-90 & G.W.Greenlees & 1971 & 9.7   & K.Matsuda & 1967 & 14.7 \\
          & R.Varner & 1986 & 16.0    & J.B.Ball & 1964 & 22.5 \\
          & R.De.Swiniarski & 1977 & 30.0    & L.N.Blumberg & 1966 & 40.0 \\
          & C.B.Fulmer & 1969 & 61.4  & H.Sakaguchi & 1982 & 65.0 \\
          & A.Nadasen & 1981 & 80.0,~135.0,~160.0 & V.Comparat & 1974 & 156.0 \\
          & E.Hagberg & 1971 & 185.0   &       &  \\
\hline
    50-Sn-120 & G.W.Greenlees & 1971 & 9.7   & R.Varner & 1986 & 16.0 \\
          & W.Makofske & 1972 & 16.0    & S.D.Wassenaar & 1989 & 20.4 \\
          & B.W.Ridley & 1964 & 30.3  & L.W.Put & 1971 & 30.4 \\
          & G.S.Mani & 1971 & 49.4  & F.E.Bertrand & 1970 & 61.5 \\
          & S.Kailas & 1984 & 104.0   & P.Schwandt & 1982 & 135.0 \\
          & V.Comparat & 1974 & 156.0   & H.Takeda & 2003 & 200.0,~250.0 \\
\hline
    82-Pb-208 & W.Makofske & 1972 & 16.0    & W.T.H Van Oers & 1974 & 21.0,~24.1,~26.3,~30.3,~35.0,\\
          &            &       &       &             &      & 45.0,~47.3 \\
          & D.W.Devins & 1962 & 30.8  & L.N.Blumberg & 1966 & 40.0 \\
          & C.B.Fulmer & 1969 & 61.4  & H.Sakaguchi & 1982 & 65.0 \\
          & A.Nadasen & 1981 & 80.0,~121.0,~160.0,~182.0 & V.Comparat & 1974 & 156.0 \\
          & C.Djalali & 1982 & 201.0   &       &  \\
    \end{tabular}%
   \end{ruledtabular}
  \label{tab6}%
\end{table*}%

\subsubsection{The proton reaction cross section}

\begin{figure}[htbp]
\centerline{\includegraphics[width = 3.5in]{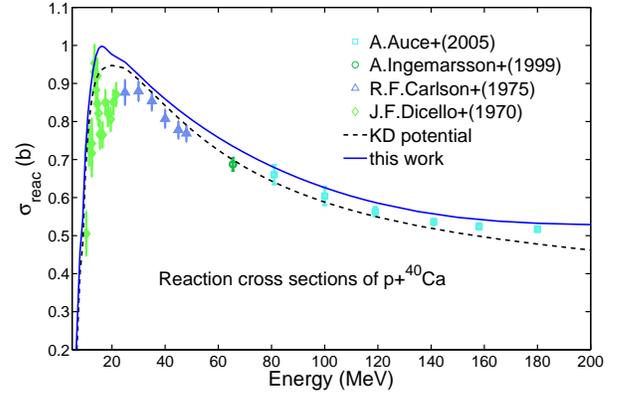}} \caption{
(color online) Comparison of predicted reaction cross section (solid line) and
experimental data (point) and KD calculation (dashed line) for $p$ + $^{40}$Ca.}
\label{fig15}
\end{figure}

\begin{figure}[htbp]
\centerline{\includegraphics[width = 3.5in]{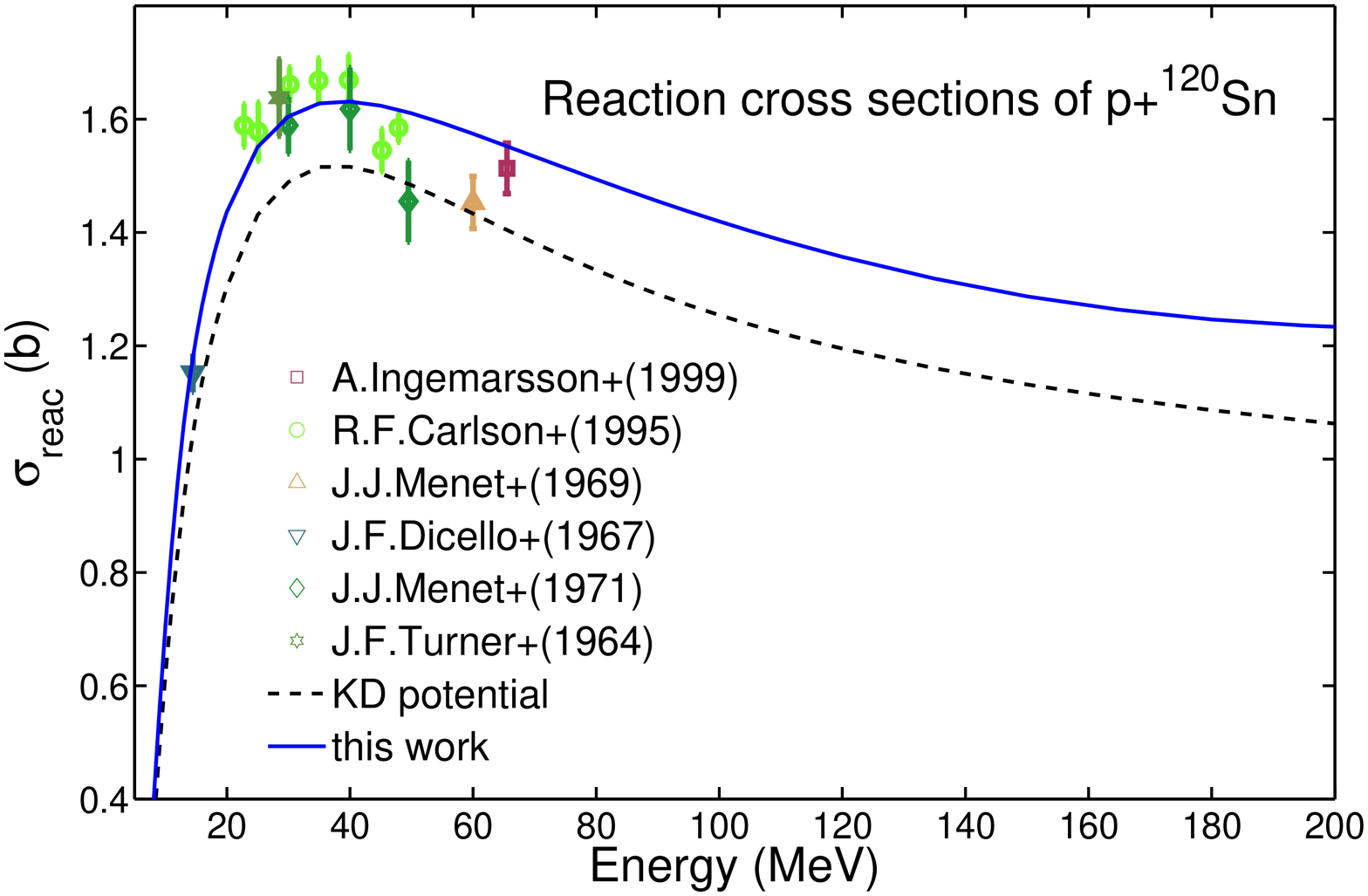}} \caption{
(color online) Comparison of predicted reaction cross section (solid line) and
experimental data (point) and KD calculation (dashed line) for $p$ + $^{120}$Sn.}
\label{fig16}
\end{figure}

\begin{figure}[htbp]
\centerline{\includegraphics[width = 3.5in]{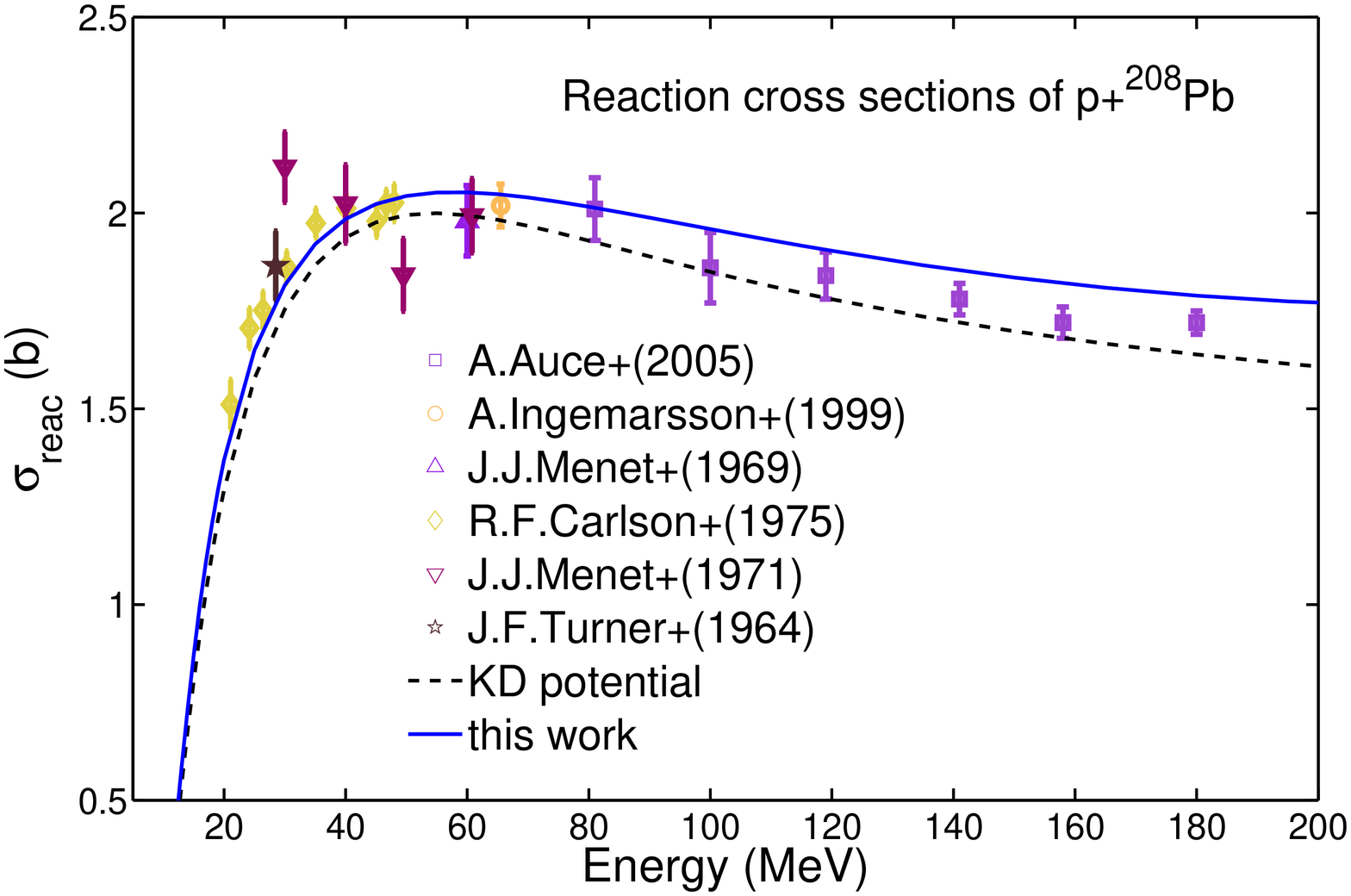}} \caption{
(color online) Comparison of predicted reaction cross section (solid line) and
experimental data (point) and KD calculation (dashed line) for $p$ + $^{208}$Pb.}
\label{fig17}
\end{figure}

It is noticed that the experimental data of proton reaction cross
sections, $\sigma_{reac}$, are much less than $\sigma_{tot}$ for
neutron both in quantity and in quality. Therefore, we also refer to
the calculated $\sigma_{reac}$ by KD potential in the process of
visual goodness-of-fit estimation. We sample the predicted
$\sigma_{reac}$ for $^{40}$Ca, $^{120}$Sn and $^{208}$Pb in Figs. \ref{fig15}, \ref{fig16} and \ref{fig17}.
It can be observed that the present calculations are good but only slightly overestimate the
experimental values in the whole energy region. In some cases, such
as $p$ + $^{120}$Sn in the lower energy region, this MOP looks
better than the global KD. After comparisons, the maximum deviation
between the predicted reaction cross sections and measurements is
less than 20\%.

\subsubsection{The elastic scattering angular distribution}

\begin{figure*}[htbp]
\centerline{\includegraphics[width = 5.0in]{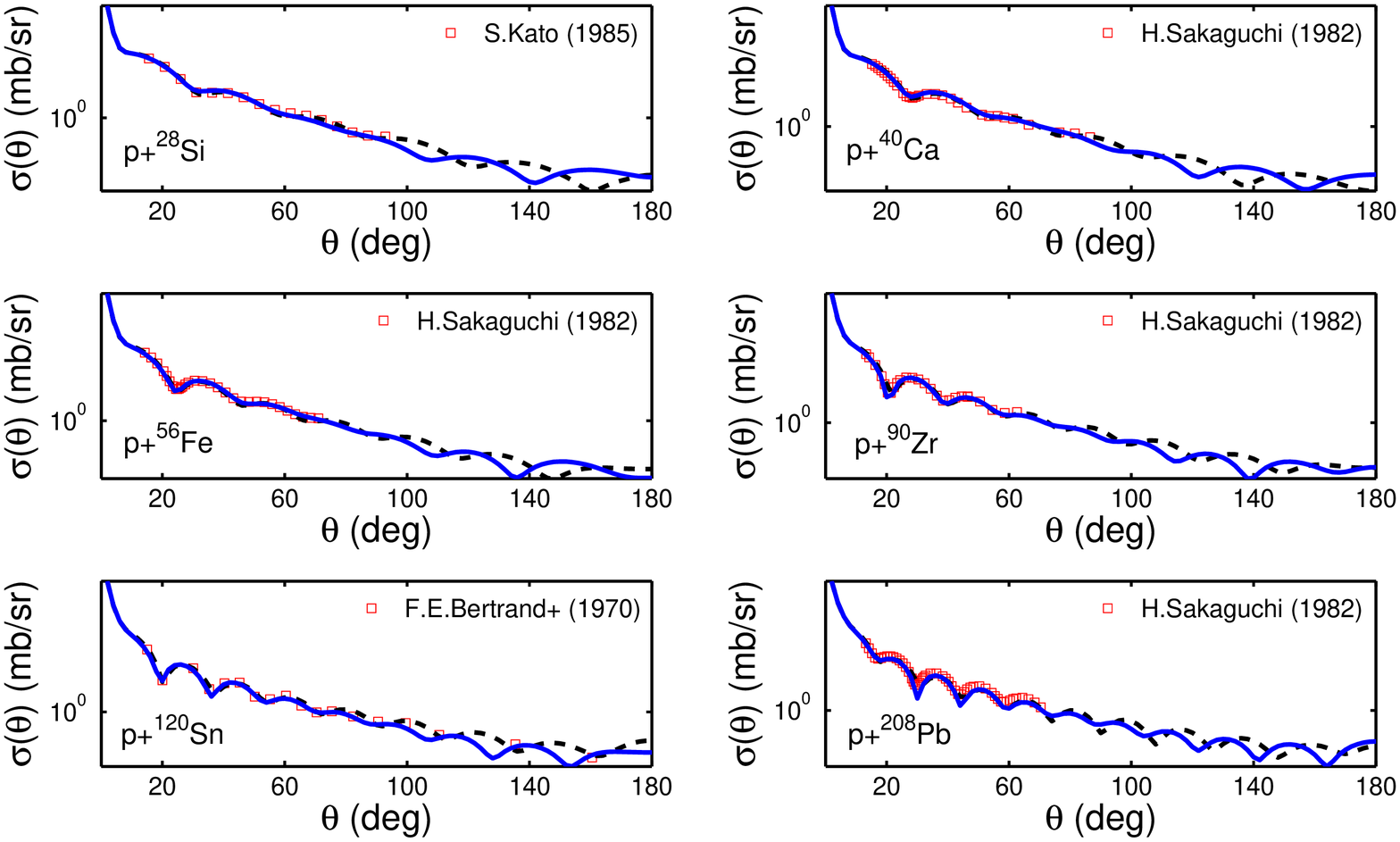}}
\caption{ (color online) Comparison of predicted angular
distribution (solid line) and experimental data (point) and KD
calculation (dashed line) for $p$ + $^{28}$Si, $^{40}$Ca, $^{56}$Fe,
$^{90}$Zr, and $^{208}$Pb at incident neutron energy 65MeV; 61.5MeV
for $p$ + $^{120}$Sn.} \label{fig18}
\end{figure*}

\begin{figure}[htbp]
\centerline{\includegraphics[width = 4.0in]{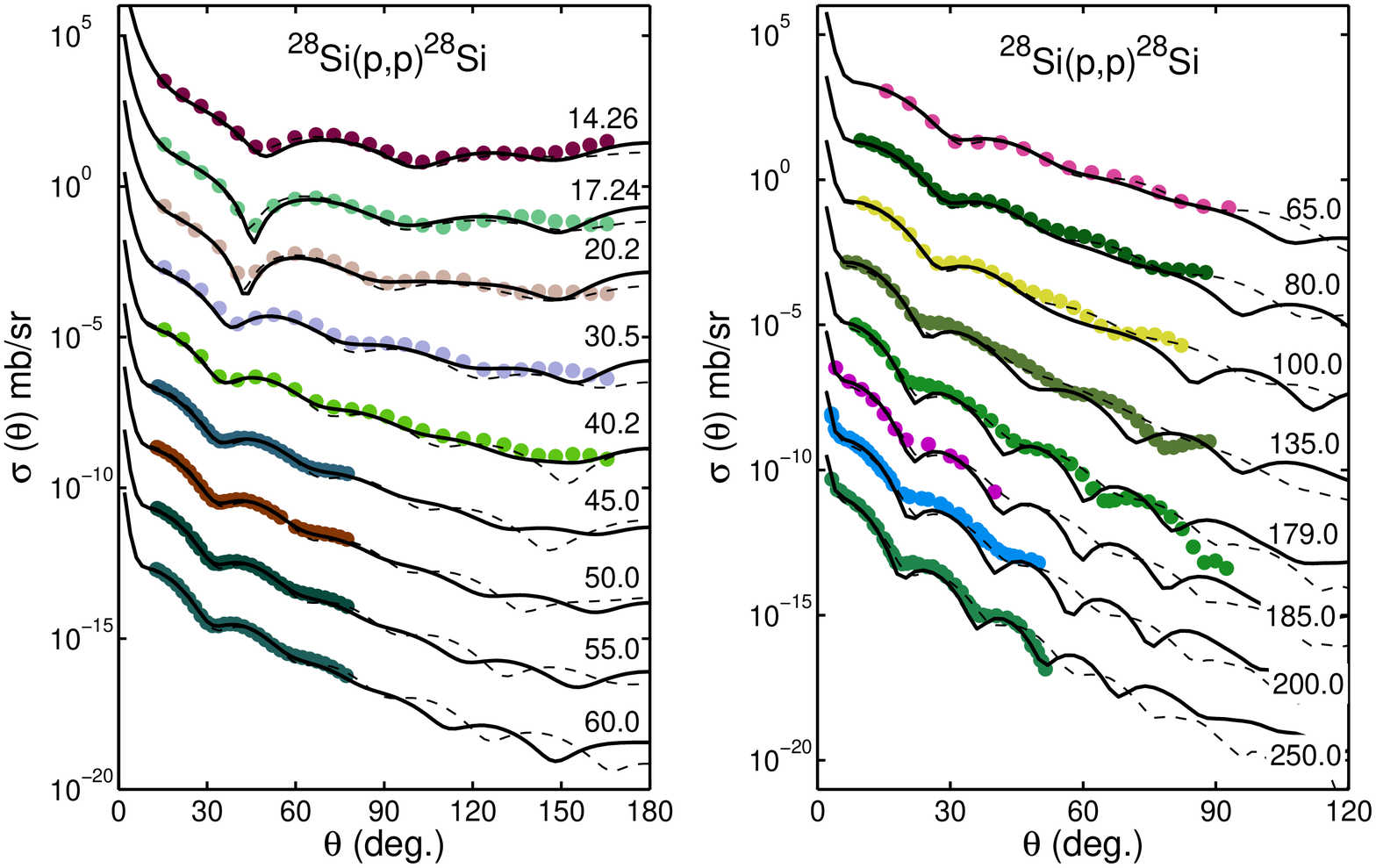}} \caption{
(color online) Comparison of predicted $d\sigma/d\Omega$ (solid line) and experimental data (point)
and KD calculation (dashed line) for $p$ + $^{28}$Si.}
\label{fig19}
\end{figure}

We collect the proton elastic scattering angular distribution just
as the case for neutron. The $d\sigma/d\Omega$ of proton scattering
from 6 nuclei, $^{28}$Si, $^{40}$Ca, $^{56}$Fe, $^{90}$Zr,
$^{120}$Sn and $^{208}$Pb, around proton incident energy at 65 MeV
are collected in Fig. \ref{fig18}. The perfect agreement between the
present calculations and experimental data displays the powerful
prediction ability of this MOP. In addition, we also condense
$d\sigma/d\Omega$ curves of various energies belonging to the same
nucleus in one figure, as in Fig. \ref{fig19}. Similarly, in these
condensed figures, the curves and data points at the top are true
values, while the others are offset by factors of 0.01, 0.0001, etc,
and incident laboratory energies are in MeV.

\textbf{Targets $^{12}$C-$^{40}$Ca:} with respect to the
differential cross section $d\sigma/d\Omega$, the resulting
$\chi^{2}/N$ of nuclei in this mass region are listed in Table
\ref{tab7}. The values for $^{12}$C and $^{28}$Si are obviously
larger than for the other nuclei. To explore the source of this
discrepancy, we focus our discussion on $^{28}$Si (see
Fig. \ref{fig19}). It is observed that the theoretical results and
the measurements are in good agreement within the entire angular
region for incident energies E$_p <$ 120 MeV. At higher energies,
however, our predictions tend to underestimate the data for the
differential cross section. This feature is the main resources to
cause the poor $\chi^2/N$.

\begin{table}
\caption{\label{tab7} The $\chi^{2}/N$ of $d\sigma/d\Omega$ for $p + ^{12}$C-$^{40}$Ca reactions}
\begin{ruledtabular}
\begin{tabular}{cccc}
 Nuclide &  Point num. of exp. &  MOP  & KD\\
\hline
 $^{12}$C & 637 & 3.70 & 0.34\\
\hline
 $^{27}$Al & 336 & 0.88 & 0.90\\
\hline
 $^{28}$Si & 513 & 4.78 & 0.56\\
\hline
 $^{40}$Ca & 682 & 0.37 & 0.22\\
\end{tabular}
\end{ruledtabular}
\end{table}

\begin{figure}[htbp]
\centerline{\includegraphics[width = 4.0in]{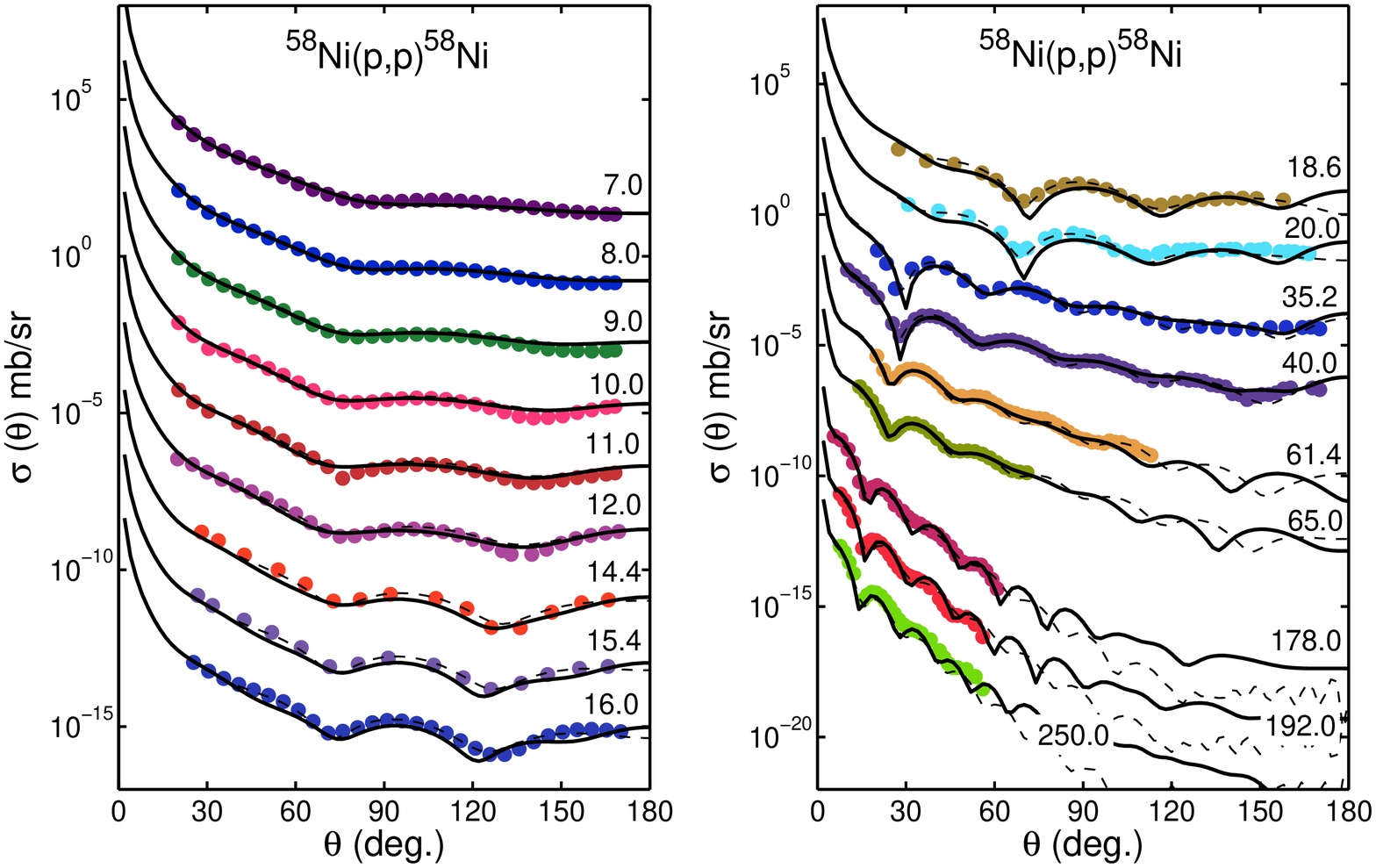}} \caption{
(color online) Comparison of predicted $d\sigma/d\Omega$ (solid
line) and experimental data (point) and KD calculation (dashed line)
for $p$ + $^{58}$Ni.} \label{fig20}
\end{figure}

\textbf{Targets $^{48}$Ti - $^{208}$Pb:} The $\chi^{2}/N$ of
$d\sigma/d\Omega$ in this target region is shown in Table
\ref{tab8}. The $\chi^{2}/N$ values show a good prediction in this
target region. Some of them are even lower than the corresponding
values by KD potential. We look through the details by considering
$d\sigma/d\Omega$ of $^{58}$Ni in Fig. \ref{fig20}. The present
predictions, measurements and phenomenological KD results are
consistent perfectly with each other in the entire energy region.

\begin{figure}[htbp]
\centerline{\includegraphics[width = 4.0in]{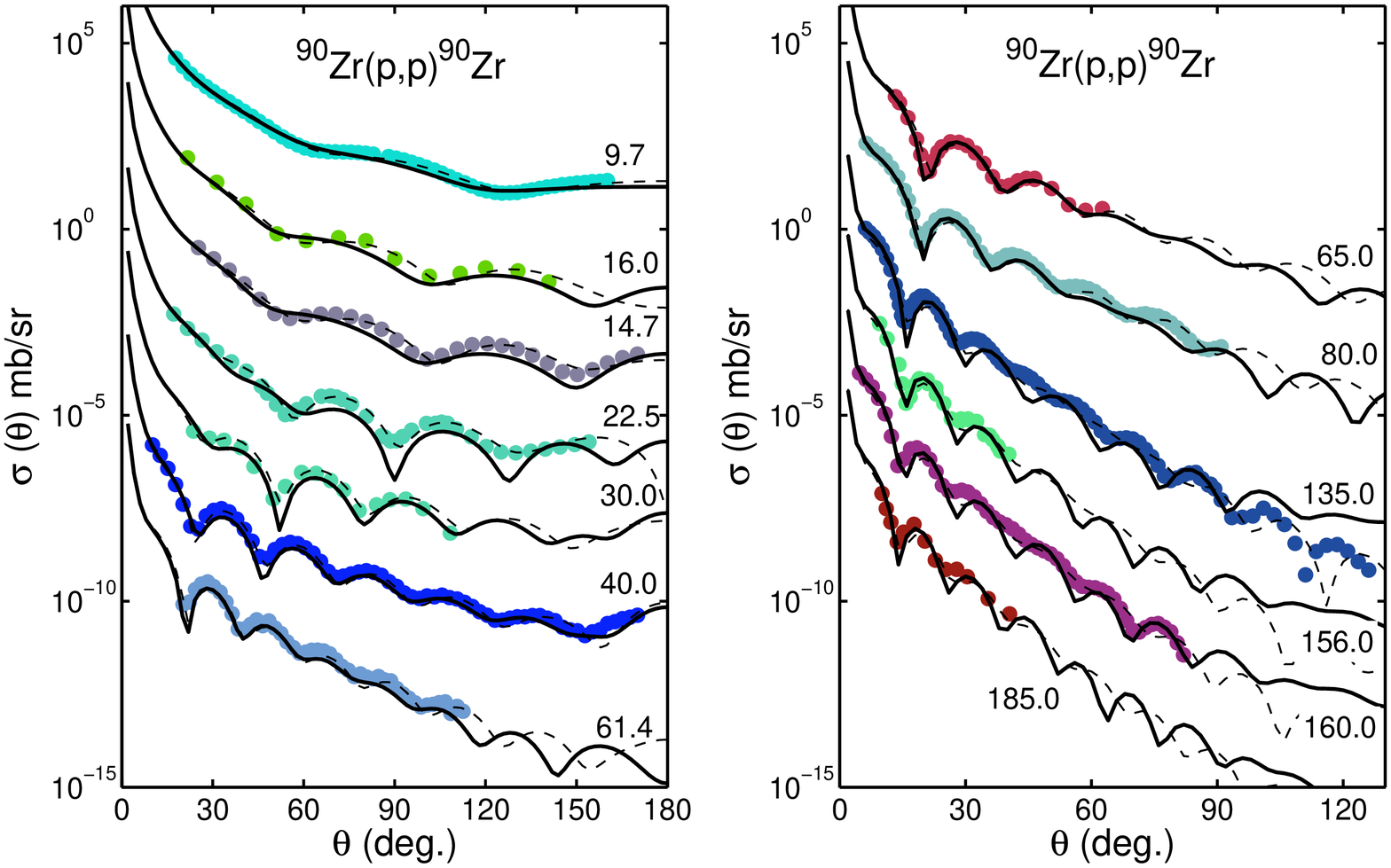}} \caption{
(color online) Comparison of predicted $d\sigma/d\Omega$ (solid
line) and experimental data (point) and KD calculation (dashed line)
for $p$ + $^{90}$Zr.} \label{fig21}
\end{figure}

\begin{figure}[htbp]
\centerline{\includegraphics[width = 4.0in]{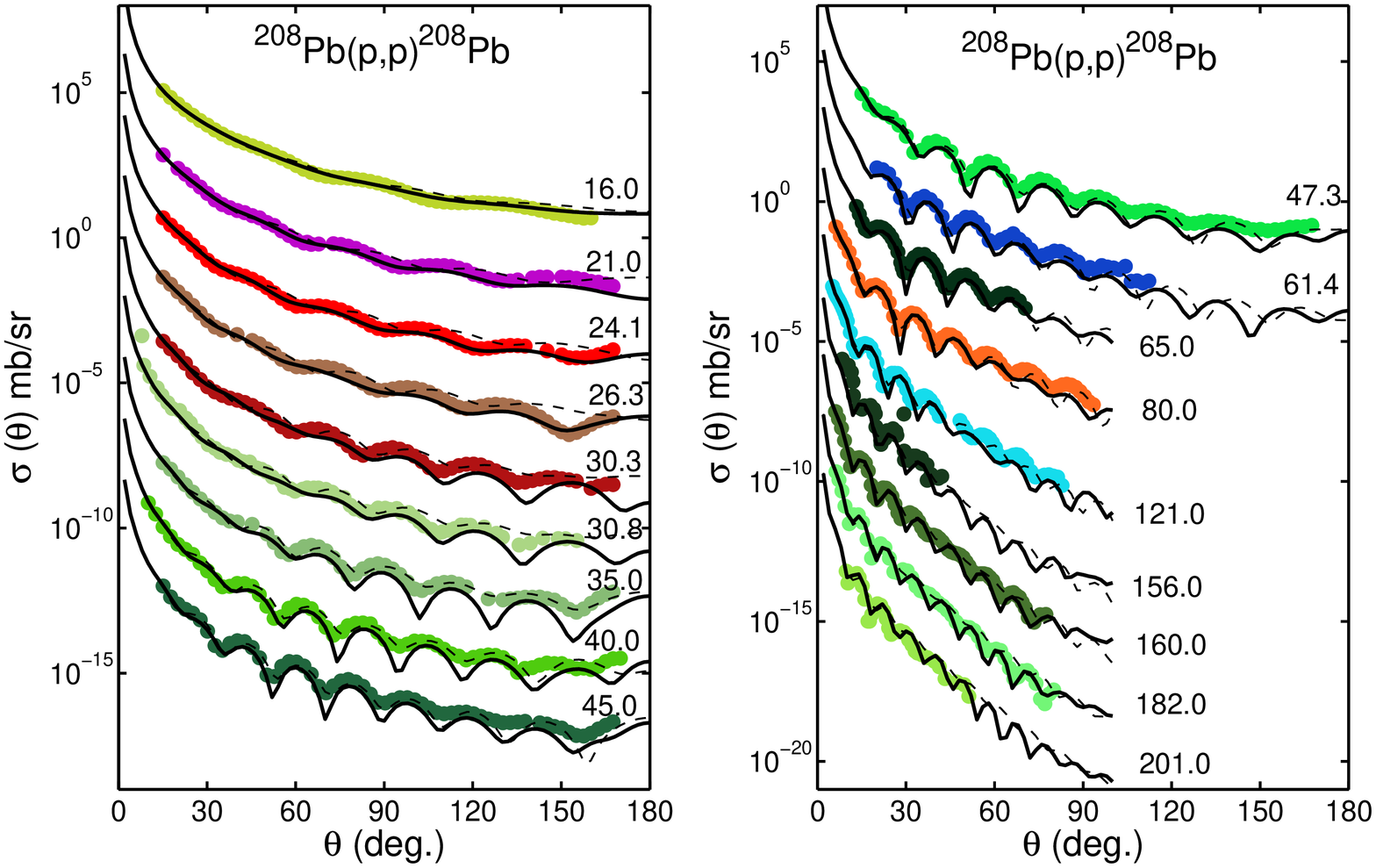}} \caption{
(color online) Comparison of predicted $d\sigma/d\Omega$ (solid
line) and experimental data (point) and KD calculation (dashed line)
for $p$ + $^{208}$Pb.} \label{fig22}
\end{figure}

$^{90}$Zr is the only example for which the $\chi^{2}/N$ is not
particularly good. Therefore we compare the calculated $d\sigma/d\Omega$ for $^{90}$Zr in
Fig.\ref{fig21}. It is found that most theoretical values are
consistent with the measurements, and the main deviations occur at
specific incident energies such as 22.5 and 135.0 MeV. In addition, a very good
performance of the MOP also occurs in the calculations for $p$ + $^{208}$Pb, as shown in Fig. \ref{fig22}.

\begin{table}
\caption{\label{tab8} The $\chi^{2}/N$ of $d\sigma/d\Omega$ for
$p+^{56}$Fe-$^{208}$Pb reactions}
\begin{ruledtabular}
\begin{tabular}{cccc}
 Nuclide &  Point num. of exp. &  MOP  & KD\\
\hline
 $^{56}$Fe & 516 & 0.16 & 0.12\\
\hline
 $^{58}$Ni & 557 & 0.15 & 0.13\\
\hline
 $^{90}$Zr & 536 & 3.61 & 0.29 \\
\hline
 $^{120}$Sn & 406 & 0.27 & 0.85 \\
\hline
 $^{208}$Pb & 1028 & 0.29 & 0.72 \\
\end{tabular}
\end{ruledtabular}
\end{table}


\subsubsection{The analyzing power and spin rotation function}

\begin{figure}[htbp]
\centerline{\includegraphics[width = 4.0in]{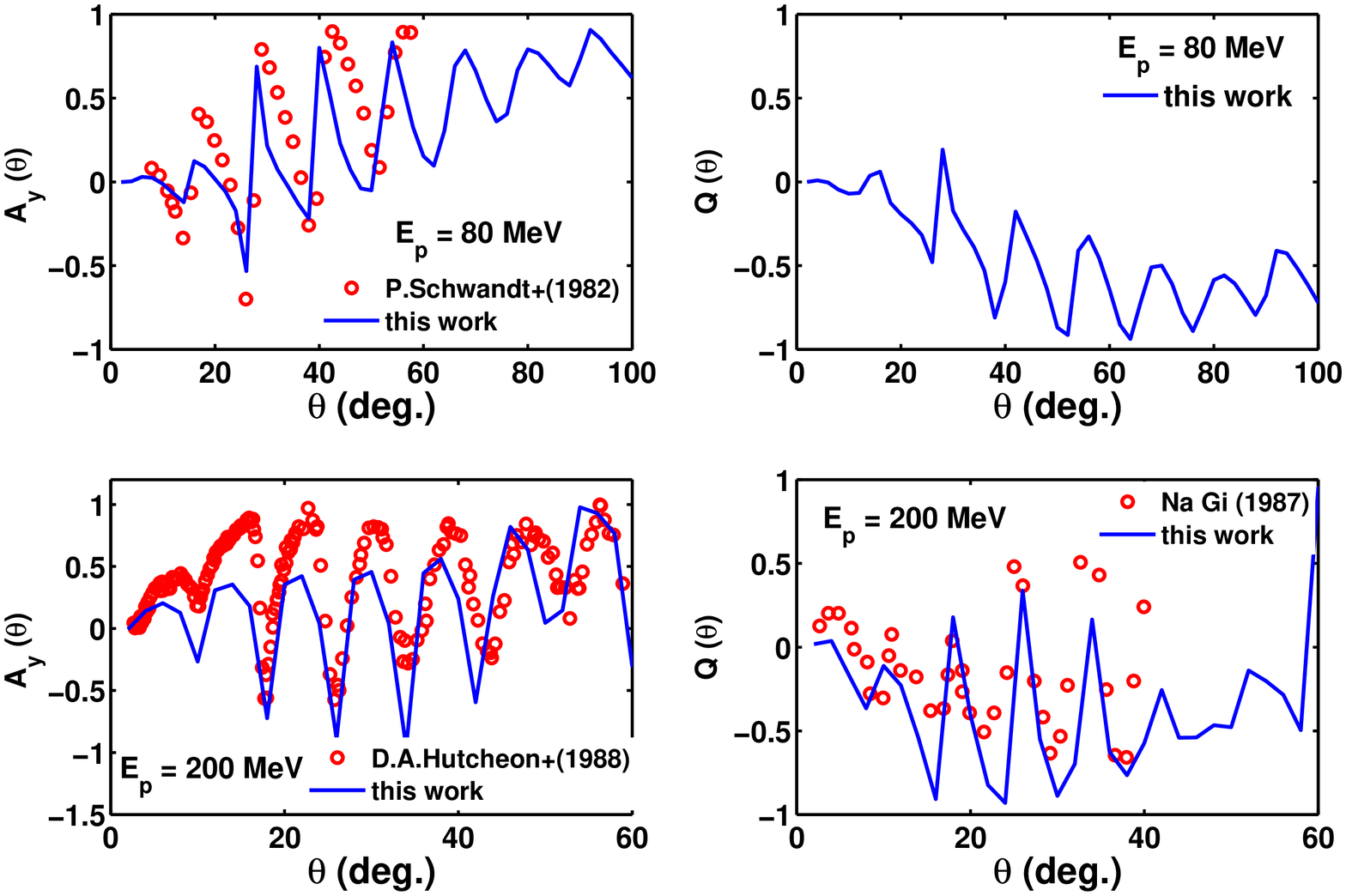}}
\caption{ (color online) Comparison of predicted A$_y$ and Q (solid
line) and experimental data (point) and KD calculation (dashed line)
for $p$+ $^{208}$Pb at $E_{p}$ = 80MeV and 200MeV. The experimental
data of A$_y$ are taken from EXFOR library, while the data of Q are
read from \cite{Ch.Elster1998}.} \label{fig23}
\end{figure}

\begin{figure}[htbp]
\centerline{\includegraphics[width = 4.0in]{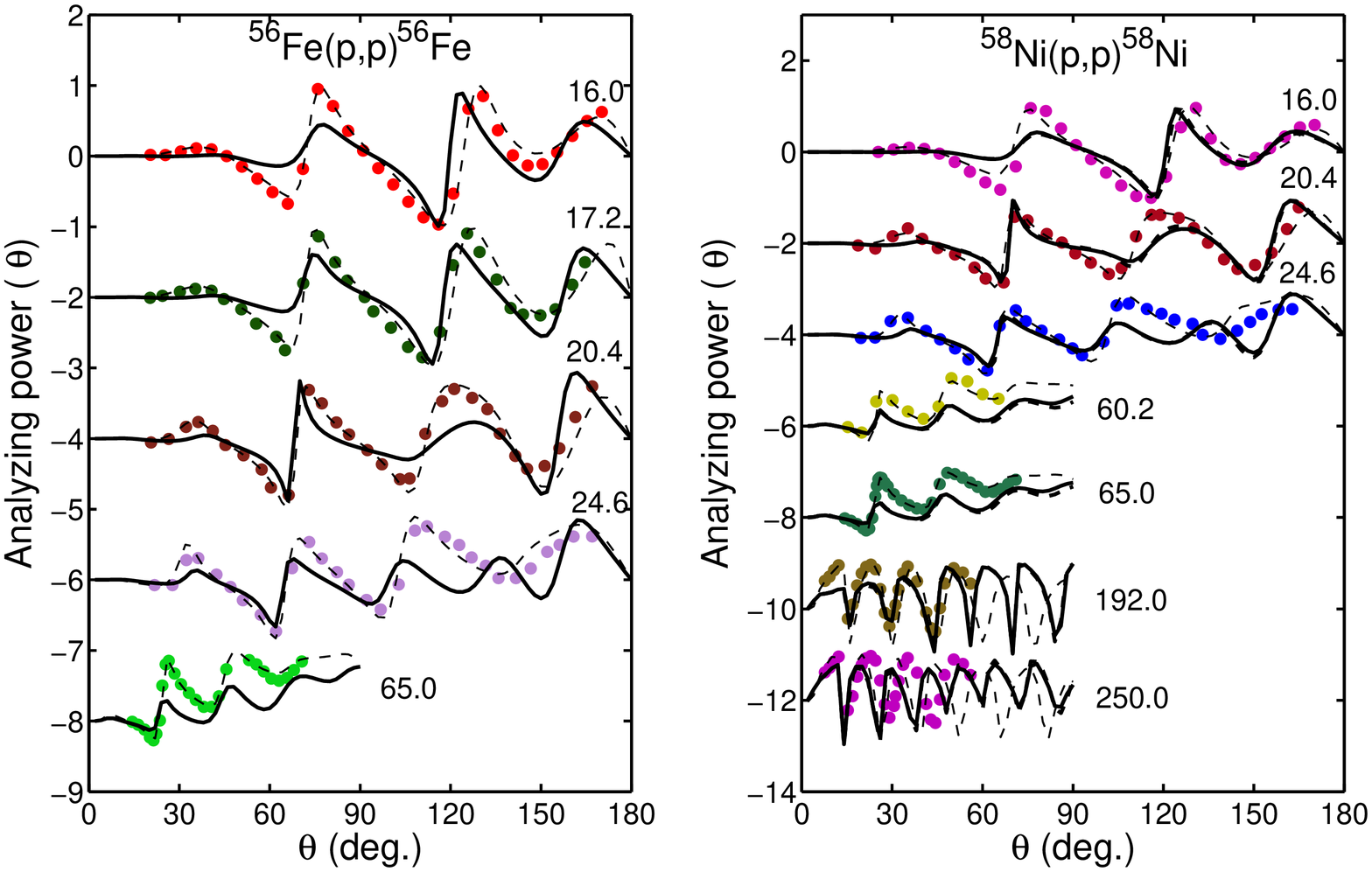}} \caption{
(color online) Comparison of predicted A$_y$ (solid line) and
experimental data (point) and KD calculation (dashed line) for $p$+
$^{56}$Fe and $^{58}$Ni. The curves and data points at the top
represent true values, the others are offset by factors of 2, 4, 8,
etc.} \label{fig24}
\end{figure}

The predicted analyzing power A$_y$($\theta$) and spin rotation
function Q($\theta$) of proton scattering from $^{208}$Pb at E$_{p}$
= 80 and 200 MeV are plotted in Fig. \ref{fig23}. The predicted
phases of A$_y$($\theta$) and Q($\theta$) look well, whereas the
amplitudes are not ideal, which remain to be improved in the future.
In addition, in order to show more results of other nuclei, we also
plot A$_y$($\theta$) of $^{56}$Fe and $^{58}$Ni in Fig. \ref{fig24},
where the applied experimental data are listed in Table \ref{tab9}.
It is shown that the amplitudes of A$_y$($\theta$) by
MOP are better around the lower energy region.

\begin{table}[htbp]
  \centering
  \caption{The $A_y$ database for proton elastic scattering from $^{56}$Fe and $^{58}$Ni.}
    \begin{ruledtabular}
    \begin{tabular}{lllllll}
    \textbf{Target} & \textbf{Author(1$^{st}$)} & \textbf{Year} & \textbf{Energy(MeV)} \\
\hline
    26-Fe-56 & R.Varner     & 1986 & 16.0    \\
             & P.J.Van.Hall & 1977 & 17.2,~20.4,~24.6 \\
             & R.De.Leo     & 1996 & 65.0  \\
\hline
    28-Ni-58 & R.Varner     & 1986 & 16.0   \\
             & P.J.Van.Hall & 1977 & 20.4,~24.6 \\
             & D.C.Kocher   & 1976 & 60.2  \\
             & H.Sakaguchi  & 1982 & 65.0,~192.0\\
             & H.Takeda     & 2003 & 250.0 \\
    \end{tabular}%
   \end{ruledtabular}
  \label{tab9}%
\end{table}%

\section{Summary}
The central attempt of this study is to present an optical model for
nucleon-nucleus scattering, which is based on a microscopic
calculation of the nucleon self-energy in nuclear matter. After
adjustment of very few parameters, the new optical potential
reproduces many nucleon scattering data for target nuclei across the
whole nuclear mass table of stable isotopes between $^{12}$C to
$^{208}$Pb at nucleon incident energies from 1 MeV to 200 MeV.
The hope is that this approach for a microscopic optical potential yields
reliable predictions also for unstable target nuclei, since an
extrapolation in this unstable region should be more reliable than a
pure phenomenological fit with more fit parameter. Of course, the
specification of a sound nuclear structure for targets, especially
for exotic nuclei remains to be progressed.

The microscopic basis of this study are Dirac Brueckner Hartree-Fock
calculations of nuclear matter using realistic forces, which have
been adjusted to describe nucleon-nucleon scattering phases. One of
the basic features of this relativistic approach is that it provides
a specific energy-dependence for the optical model and also predicts
a spin-orbit term without the need to introduce any additional
parameters (see e.g. \cite{dalrev}).

The complex isospin dependent self-energies are extracted from the
DBHF approach with projection techniques using the Bonn-B bare $NN$
interaction. The MOP with Bonn A has  also been  tested and the
results show that the prediction of the scattering for finite nuclei
is not very sensitive to the choice of a realistic nucleon force,
Bonn A or Bonn B. Therefore, Bonn-B has been adopted following our
pilot study \cite{RRXU2012}. The present MOP is very strictly built
on the DBHF calculations in nuclear matter at $\rho>0.08$ fm$^{-3}$
by means of the improved local density approximation. For the
purpose of describing the observables of scattering, we construct
the optimization method according to the annealing algorithm. The
range factors in ILDA and the scalar and vector potentials below
0.08fm$^{-3}$ are extracted using this method from the experimental
data of $^{40}$Ca as an example for isospin symmetric nuclei and
$^{208}$Pb for isospin asymmetric nuclei. Then they are applied for
many nuclei and energy regions. Good predictions for most nuclei are
achieved by the resulting  MOP only with the free parameter $t$. The
results of the MOP are of a quality, which is comparable to the
widely used phenomenological Koning-Delaroche (KD) global potential.
A bit of imperfection still exists in both $n$+A and $p$+A systems
for specific target nuclei.

Certainly, it is impossible to depict all quantities in a perfect
way within the spherical nuclear OM and the present MOP, especially
for the strongly deformed nuclei around rare earth and actinide.
Also it should be kept in mind that the MOP is based on a
microscopic study of nuclear matter. Therefore all features, which
are related to surface excitation, e.g. the particle-vibration
coupling, are not explicitly taken into account. Such features shall
be included in future studies. Moreover we plan to make the present
MOP available in form of an interactive web-based application.

\begin{acknowledgments}
We thank Prof. S. Hilaire for kindly supplying us all the density
distributions of finite nuclei calculated in the
Hartree-Fock-Bogoliubov (HFB) approach with Gogny D1S force. R. R.
Xu thanks Prof. Q. B. Shen and Y. L. Han for helpful discussion on
the code APMN and physics. This work has been supported by the
National Basic Research Program of China under Grant No 2013CB834404
and the National Natural Science Foundation of China (Grant Nos.
11305270, 11275018); the Deutsche Forschungsgemeinschaft (DFG) under
contract no. Mu 705/10-1.
\end{acknowledgments}

\end{CJK*}
\end{document}